\documentclass[showpacs,pre,aps,superscriptaddress,twocolumn,floatfix]{revtex4}

\usepackage{lineno,hyperref,mathtools}
\usepackage[usenames,dvipsnames]{xcolor}
\usepackage{amsmath}
\usepackage{amssymb,mathrsfs}
\usepackage{graphicx}

\newcommand{\prt}{\partial}
\newcommand{\la}{\lambda}
\newcommand{\vphi}{\varphi}

\newcommand{\sn}{\mathrm{sn}}
\newcommand{\cn}{\mathrm{cn}}
\newcommand{\ts}{\tilde{s}}
\newcommand{\n}[1]{N_{#1}}
\usepackage{hyperref}

\begin{document}

\title{Solution of the Riemann problem for polarization waves in a
  two-component Bose-Einstein condensate}

\author{S. K. Ivanov} \affiliation{Institute of Spectroscopy, Russian
    Academy of Sciences, Troitsk, Moscow, 108840, Russia}
\affiliation{Moscow Institute of Physics and Technology, Institutsky
    lane 9, Dolgoprudny, Moscow region, 141700, Russia}

\author{A. M. Kamchatnov}
\affiliation{Institute of Spectroscopy, Russian Academy
  of Sciences, Troitsk, Moscow, 108840, Russia}
\affiliation{Moscow Institute of Physics and Technology, Institutsky
    lane 9, Dolgoprudny, Moscow region, 141700, Russia}

\author{T. Congy}\author{N. Pavloff}
\affiliation{LPTMS, CNRS, Univ. Paris-Sud, Universit\'e
Paris-Saclay, 91405 Orsay, France}

\begin{abstract}
  We provide a classification of the possible flow of two-component
  Bose-Einstein condensates evolving from initially discontinuous
  profiles. We consider the situation where the dynamics can be
  reduced to the consideration of a single polarization mode (also
  denoted as ``magnetic excitation'') obeying a system of equations
  equivalent to the Landau-Lifshitz equation for an easy-plane
  ferro-magnet. We present the full set of one-phase periodic
  solutions. The corresponding Whitham modulation equations are
  obtained together with formulas connecting their solutions with the
  Riemann invariants of the modulation equations. The problem is not
  genuinely nonlinear, and this results in a non-single-valued mapping
  of the solutions of the Whitham equations with physical wave
  patterns as well as to the appearance of new elements --- contact
  dispersive shock waves --- that are absent in more standard,
  genuinely nonlinear situations. Our analytic results are confirmed
  by numerical simulations.
\end{abstract}

\pacs{67.85.Fg,47.35.Fg,75.78.-n}
\maketitle

\section{Introduction}\label{sec1}

The first experimental realizations of Bose-Einstein condensation
(BEC) of single species ultracold atomic vapors \cite{And95,Dav95}
were soon followed by their multi-component counterparts \cite{Mya97}
which appeared to be nontrivial extensions of the previous ones, the
dynamical and nonlinear aspects of phase separation revealing
particularly rich \cite{Ste98,Hal98}.  Over the years, numerous
studies have been devoted to theoretical and experimental
investigations associated with these specific features, namely
nonlinearity and dynamics in multi-component BECs, see, e.g., the
reviews \cite{Kaw12,Sta13} and chapters in the books
\cite{PeSm,PiSt,Kev08}.

The specific physical ingredients of this body
of research are the (intra- and inter-species) interactions, the negligible
viscosity and the large dispersive effects.
Another important aspect is the different degrees of freedom
associated with the different types of motion of the components. For two
component systems one can schematically separate global in-phase
motion --- associated with density fluctuations --- from out-of-phase
motion, associated with a ``polarization'' or ``magnetic'' degree of
freedom. This appealing classification of the dynamical behaviors of
the system is however oversimplified: in many instances, a clean
separation between these idealized types of excitation is not
possible, even at the perturbative level (see, e.g., the discussion in
\cite{kklp-14}). However, a recent theoretical breakthrough has been
made in Ref.~\cite{qps-16} where it has been shown that for stable two
component mixtures close to the immiscibility region, the density and
magnetization degrees of freedom decouple, even at the nonlinear
level. The polarization sector is particularly interesting, new solitons
have been first identified in Ref.~\cite{qps-16} and a rich variety of
nonlinear excitations rapidly followed \cite{Con16}: cnoidal waves,
nonlinear trigonometric waves, algebraic solitons. The interest of
these studies is not uniquely theoretical: the regime of parameters
for which the dynamics of polarization excitations decouples from that
of density excitations corresponds to systems of experimental
interest; for instance, it is exactly realized in the mixture of the
two hyperfine states $|F = 1, m_F = \pm 1\rangle $ of $^{23}$Na
\cite{Bien16}, and, to a good approximation, in the mixture of
hyperfine states of $^{87}$Rb considered in Refs.~\cite{Ham11} ($|1,
1\rangle$ and $|2, 2\rangle $) and \cite{Dan16} ($|1, -1\rangle$ and
$|1, 0\rangle $ or $|1, -1\rangle$ and $|2, -2\rangle $).

An investigation of the one-dimensional Riemann problem for
polarization excitations was started in Ref.~\cite{Con16}, which was
motivated by the study of two-species counterflow considered in
Ref.~\cite{Ham11}: an initial value problem has been considered,
consisting, for each component, in piece-wise constant initial
(relative) density and velocity, with a single discontinuity. The
importance of this type of problems lies in the facts that, first,
their solution involves characteristic wave patterns arising in the
space-time evolution of quite general initial pulses, and, second, a
number of real physical situations can be reduced to the discussion of
the dynamics of initial discontinuities. The interest of this
so-called ``Riemann problem'' was first realized in the framework of
compressible fluid dynamics, where the well-known viscous shocks play
a key role in the classification of evolutions of initial
discontinuities (see, e.g., Ref.~\cite{LL-6}).  Extension of this
approach to systems where dispersion effects play a dominant role ---
instead of viscous ones --- started with the ground breaking work of
Gurevich and Pitaevskii \cite{gp-73} for the Korteweg-de Vries (KdV)
equation, in which dispersive shock waves (DSWs) were approximated by
nonlinear modulated waves whose evolution were described by means of
Whitham theory of modulations \cite{whitham-65,Whithbook}.  The theory
of DSWs has been much developed since and has found numerous different
applications (see, e.g., a recent review \cite{eh-16} and references
therein). In particular, the classification of the space-time
evolution of initial discontinuities was established for waves whose
dynamics is described by the nonlinear Schr\"odinger (NLS) equation
\cite{gk-87,El95} and by the Kaup-Boussinesq (KB) equation
\cite{egp-01,KB17}.  In all these cases (KdV, NLS and KB), the
evolution of the DSW is governed by the dynamical Whitham equations
for the so called ``Riemann invariants'' \cite{LL-6} who have a
one-to-one correspondence with relevant physical variables. However,
the problem becomes much more complicated when this mapping is
multi-valued, even when the Whitham equations can be represented in a
diagonal form. For instance, new types of structures appear in such
systems, as was indicated in Ref.~\cite{marchant} for the case of the
modified KdV (mKdV) equation. The full solution of the Riemann problem
for the Gardner equation (related with the modified KdV equation) was
given in Ref.~\cite{Kam12} and this solution was adapted to the mKdV
case in Ref.~\cite{El17}. These examples refer to a unidirectional
wave propagation described by a single nonlinear wave equation.
However, similar complicated wave structures were discussed in
Ref.~\cite{ep-11} for the non-integrable Miyata-Camassa-Choi equation
describing two-directional propagation of two-layer shallow water
waves.

The case considered in Ref.~\cite{Con16} combines two difficulties:
(i) as in the problems studied in Refs.~\cite{marchant,Kam12,El17}, it
corresponds to a situation where the dispersionless Riemann invariants
are non-monotonously dependent on the physical variables, that is, the
problem is not {\it genuinely nonlinear} (see, e.g., \cite{lax}) and,
(ii) as in Ref.~\cite{ep-11}, it corresponds to a two-directional wave
propagation described by a system of two nonlinear equations.
To avoid too many complications, the study of Ref.~\cite{Con16} was
initially restricted to a region of parameters where the dependence of
the Riemann invariants on the physical variables remains monotonous,
that is, the problem considered was actually genuinely nonlinear. In
the present paper we extend this study and give the full solution of
the Riemann problem for the space-time evolution of polarization waves
in a two-component BEC. Our approach is based on the remark made in
Ref.~\cite{Con16} that, for the regime of parameters identified in
Ref.~\cite{qps-16}, nonlinear polarization waves can be described by
the dissipationless Landau-Lifshitz (LL) equation with uniaxial
easy-plane anisotropy \cite{ll-1935,Lif78}. The exact integrability of
this equation --- which belongs to the Ablowitz-Kaup-Newell-Segur
hierarchy --- makes it possible to develop a Whitham modulational theory
(Sec.~\ref{sec4}) for describing configurations where nonlinear waves
are slowly modulated, as observed in dispersive shocks. This will
permit us to formulate a principle of classification valid for all the
numerous wave patterns arising from the evolution of initial
discontinuities.

An interesting aspect of the present work is its relevance to systems
pertaining to widely different domains in physics. Configurations
similar to the ones studied in the present work can be investigated
in neighboring fields such as nonlinear fiber optics and also
exciton-polariton condensed systems. But the physical ingredients
characterizing the phenomena we are interested in --- nonlinearity,
weak dissipation, dispersion in a multi-component system --- are also
encountered in quite different settings.  As a result, the solution of
the Riemann problem we give in the present work is also relevant to
fluid mechanics \cite{Lacombe65,Ovs79,cavanie,sq-93,Lia00} and
to the nonlinear magnetization dynamics of anisotropic ferromagnets
\cite{Kos90,Iac17b,Iac17}.

The paper is organized as follows: the model and the relevant
dynamical equations are presented in section \ref{sec2}. The exact
integrability of the easy plane Landau-Lifshitz equations is used in
Sec.~\ref{sec3} for writing its explicit one-phase solutions,
determining the corresponding Riemann invariants, and writing the
Whitham modulational equations. The full classification of
the solutions of the Riemann problem is presented in Section~\ref{sec5}
in terms of the combination of specific wave patterns,
which we denote as ``building blocks'' or ``key elements'' which are
first analyzed in Sec.~\ref{sec4}. Finally, we present our conclusions
in Sec.~\ref{sec.conclu}.

\section{The model}\label{sec2}

We consider a one-dimensional system consisting in an elongated two
component BEC described by the order parameters $\psi_\uparrow(x,t)$
and $\psi_\downarrow(x,t)$. The dynamics of the system is described by
two coupled Gross-Pitaevskii equations:
\begin{equation}\label{gp}
\left({\rm i}\hbar \partial_t +\frac{\hbar^2 \partial_x^2}{2 m}\right)
\begin{pmatrix}\psi_\uparrow\\ \psi_\downarrow\end{pmatrix}=
\begin{pmatrix}g_{\uparrow\uparrow}|\psi_{\uparrow}|^2 &
g_{\uparrow\downarrow}\psi^*_\downarrow\psi_\uparrow\\
g_{\uparrow\downarrow}\psi^*_\uparrow\psi_\downarrow &
g_{\downarrow\downarrow}|\psi_{\downarrow}|^2
\end{pmatrix}
\begin{pmatrix}\psi_\uparrow\\ \psi_\downarrow\end{pmatrix}
\; ,
\end{equation}
where $g_{\uparrow\uparrow}$ and $g_{\downarrow\downarrow}$ are the
intra-species nonlinear constants; $g_{\uparrow\downarrow}$ is the
interspecies one. We consider the limit where
$g_{\uparrow\uparrow}\approx g_{\downarrow\downarrow}$ and denote as
$g$ their common value (the situation where these two constants are
not exactly equal is treated in Ref.~\cite{kklp-14}). We denote as
$\delta g$ the difference $g-g_{\uparrow\downarrow}$ and consider the
situation
\begin{equation}\label{misc}
0 < \delta g \ll g\; .
\end{equation}
The left condition is the mean-field miscibility condition of the two
species (see, e.g., Refs.~\cite{PeSm,PiSt}). The right condition
implies that the three interaction constants are close to each other
and that the system is close to the region of immiscibility. As
discussed in Refs.~\cite{qps-16,Con16}, in this situation the density
and magnetic degrees of freedom effectively decouple.

The spinor wave function is parameterized
as \cite{Son02}
\begin{equation}\label{son-step}
\begin{pmatrix}\psi_\uparrow\\ \psi_\downarrow\end{pmatrix}=
\sqrt{\rho}\, e^{{\rm i}\Phi/2} \, \Xi \; , \quad\mbox{where}\quad
\Xi=
\begin{pmatrix}
\cos\theta \, e^{-{\rm i}\phi/2} \\
\sin\theta \, e^{{\rm i}\phi/2}
\end{pmatrix}\; .
\end{equation}
In this expression $\rho(x,t)$ is the total density and $\theta(x,t)$
governs the relative densities of the two components:
$\rho_\uparrow(x,t)=|\psi_\uparrow|^2=\tfrac12\, \rho\, (1+\cos\theta)$ and
$\rho_\downarrow(x,t)=|\psi_\downarrow|^2=\tfrac12\, \rho \, (1-\cos\theta)$.
$\Phi(x,t)$ and
$\phi(x,t)$ are potentials for the velocity fields $v_\uparrow$ and
$v_\downarrow$ of the two components, namely,
\begin{equation}
 v_{\uparrow}(x,t)=\frac{\hbar}{2m}(\Phi_x - \phi_x)\; ,\quad
v_{\downarrow}(x,t)=\frac{\hbar}{2m}(\Phi_x+ \phi_x)\; .
\end{equation}
The small perturbations of a uniform BEC of total density $\rho_0$ with
equal fractions of the two components ($\theta=\pi/2$) correspond to
total density fluctuations which propagate with velocity $c_d=
[\rho_0(g-\delta g/2)/m]^{1/2}$ and polarization excitations with
velocity $c_p=(\rho_0\delta g/2m)^{1/2}$. In the limit \eqref{misc}
these two velocities are widely different. As a result, even an
initial state consisting of a mixture of density and polarization
fluctuations rapidly separates into density perturbations propagating
at large velocity $c_d$ away from a region where only polarization
excitations take place. For considering these excitations, it is
appropriate to re-scale the lengths in units of the polarization healing
length $\xi_p=\hbar/(2 m \rho_0 \delta g)^{1/2}$ and time in units of
$\tau_p=\xi_p/c_p$. Once this is done, it has been shown in
\cite{Con16} that the dynamics of the polarization excitations is
accounted for by the following system of coupled equations
\begin{equation}\label{syst1}
\begin{split}
& \theta_t+2\,\theta_x\, \phi_x\cos\theta
+\phi_{xx}\sin\theta=0,\\
& \phi_t-\cos\theta(1-\phi_x^2)-\frac{\theta_{xx}}{\sin\theta}=0.
\end{split}
\end{equation}
The other fields are fixed by the conditions $\rho(x,t)=\rho_0$ and
$(\Phi_x - \phi_x\cos\theta)_x = 0$. Introducing the effective spin
($\sigma_x$, $\sigma_y$ and $\sigma_z$ are the Pauli matrices)
\begin{equation}\label{spin}
  \mathbf{S}=\Xi^\dagger\, \boldsymbol{\sigma}\,  \Xi
=\begin{pmatrix}
\sin\theta\cos\phi\\
\sin\theta\sin\phi\\
\cos\theta
\end{pmatrix} \; ,
\end{equation}
and the magnetization $\mathbf{M}=-\mathbf{S}$, one can easily verify that
the system of equations \eqref{syst1} is equivalent to the
dissipationless Landau-Lifshitz equation for an easy plane ferromagnet:
\begin{equation}\label{syst2}
\partial_t \mathbf{M} = \mathbf{H}_{\rm eff}\wedge \mathbf{M} \; ,
\quad\mbox{where}\quad
\mathbf{H}_{\rm eff}=\partial_x^2
\mathbf{M} - M_z \, \mathbf{e}_z \; ,
\end{equation}
$\mathbf{e}_z$ being a unit vector of the $z$-axis. We have found that
this form of the equations of motion is particularly appropriate for
numerical simulations. The reason is that, contrarily to the systems
\eqref{syst1} (and \eqref{syst3}, see below), it does not involve small
denominators when the density of one of the components gets very
small. The other interesting feature of this system is that the
anisotropic Landau-Lifshitz system \eqref{syst2} is integrable by the
inverse scattering transform method and the corresponding Lax pair is
known (see, e.g., \cite{Bor78,Bor88}). This result has been used in
Ref. \cite{Kam92} to derive periodic solutions for ferromagnets with an
easy-axis anisotropy, and we shall adapt here this approach to the
easy-plane case \eqref{syst2}.

For future convenience, we introduce a third version of \eqref{syst1}:
let us define the quantities
$w(x,t)=\cos\theta=S_z=-M_z=(\rho_\uparrow-\rho_\downarrow)/\rho_0$ describing
the variations of the relative density, and
$v(x,t)=\phi_x=(v_\downarrow - v_\uparrow)/(2c_p)$ which represents
the non-dimensional relative velocity.  In terms of these two fields
the equations of motion read
\begin{equation}
\begin{split}\label{syst3}
& w_t-[(1-w^2)v]_x=0\; ,\\
& v_t-[(1-v^2)w]_x+ \left[\frac1{\sqrt{1-w^2}}
\left(\frac{w_x}{\sqrt{1-w^2}}\right)_x\right]_x=0 \; .
\end{split}
\end{equation}
Before embarking to the study of nonlinear phenomena, it is
interesting to briefly consider linear perturbations of a stationary
configuration: let a uniform background be characterized by a
relative density $w_0$ and a relative velocity $v_0$. Small
perturbations of the type
\begin{equation}\nonumber
w=w_0+w'(x,t)\; ,\; v=v_0+v'(x,t)\; , \;\mbox{with}\;
|v'|\, , |w'| \ll 1
\end{equation}
can be sought under the form of plane waves with wave vector $k$ and
angular frequency $\omega$. Linearizing the system \eqref{syst3} one
obtains the following dispersion relation:
\begin{equation}\label{eq15}
\omega=\left(2w_0v_0\pm\sqrt{(1-w_0^2)(1-v_0^2)+k^2}\,\right)k\; .
\end{equation}
By definition we always have $|w_0|=|\cos\theta_0| \leq 1$, however
$v_0$ can have any value, and for $|v_0|>1$ the frequency $\omega$ is
complex for small enough wavevectors $k$. This implies a long
wavelength modulational instability of a system with large relative
velocity of the two components, more precisely for a background
relative velocity $v_\downarrow-v_\uparrow$ larger than $2 c_p$. This
mechanism of instability has been first theoretically studied in
Ref.~\cite{Law2001}.

In what follows, we consider the dynamically stable situation where
$|v_0|<1$. In this case the large wavelength limit of the dispersion
relation \eqref{eq15} corresponds to waves propagating with the
``polarization/magnetization sound velocity''
\begin{equation}\label{vsound}
c_{\pm} =2w_0v_0\pm\sqrt{(1-w_0^2)(1-v_0^2)} \; .
\end{equation}
For a uniform system in which both components have equal densities
($w_0=0$) and no relative velocity ($v_0=0$) one gets $c_{\pm}=\pm 1$,
i.e., going back to dimensional quantities, the speed of the magnetic
sound is $\pm c_p$ as expected. We note that the $+$ sign ($-$ sign)
in expression \eqref{vsound} corresponds to polarization excitations
propagating to the right (to the left) with respect to the background
in the reference frame in which the total flux of the condensate is
zero.

\subsection{Limiting regimes}

For some specific values of the field variables, the anisotropic
Landau-Lifshitz system \eqref{syst2} can be approximated by simpler
nonlinear models. In the present subsection we consider two limiting
cases: the nonlinear Schr\"odinger equation (Sec.~\ref{NLS0}) and the
Kaup-Boussinesq system (Sec.~\ref{KB0}).  These limiting regimes will
be used in Secs.~\ref{sec5a} and \ref{sec5b} to help classifying the
large number of different solutions of the Riemann problem.

\subsubsection{Nonlinear Schr\"odinger regime}\label{NLS0}

In the regime where $w(x,t)$ is close to unity and $v(x,t)$ is small,
defining $w'(x,1)=1-w(x,t)$ one can rewrite the system \eqref{syst3}
keeping only terms up to second order in the small quantities $v$
and $w'$:
\begin{equation}\label{appro-sys1}
\begin{split}
w'_t + 2 (w'v)_x= & 0 \; ,\\
v_t + 2 v v_x + w'_x +
\left[\frac{w'^2_x}{4 w'^2} -\frac{w'_{xx}}{2 w'}\right]_x= & 0
\; .
\end{split}
\end{equation}
Defining $n=w'/2$ and changing variable to $T=2 t$, the system
\eqref{appro-sys1} can be cast in the form
\begin{equation}\label{NLS1}
\begin{split}
n_T + (n v)_x= & 0 \; ,\\
v_T + v v_x + n_x +
\left[\frac{n^2_x}{8 n^2} -\frac{n_{xx}}{4 n}\right]_x= & 0
\; ,\end{split}
\end{equation}
which is the hydrodynamic form of the defocusing nonlinear
Schr\"odinger equation
\begin{equation}\label{NLS2}
{\rm i}\,\psi_T = - \tfrac{1}{2}\psi_{xx} + |\psi|^2\psi\; .
\end{equation}
The system \eqref{NLS1} is obtained from the standard form \eqref{NLS2}
by means of the Madelung
transform ($n=|\psi|^2$ and $v=(\mbox{arg}\,\psi)_x$).
We note that a similar approximation is also valid for
$w(x,t)$ close to $-1$ and small $v(x,t)$.

\subsubsection{Kaup-Boussinesq regime}\label{KB0}

In the regime where $v(x,t)$ is close to unity and $w(x,t)$ is small,
defining $v'(x,t)=1-v(x,t)$, one can rewrite the system
\eqref{syst3} keeping only terms up to second order in the small
quantities $v'$ and $w$:
\begin{equation}\label{appro-sys2}
\begin{split}
w_t+2w w_x + v'_x= & 0 \; , \\
v'_t+ 2(v' w)_x-w_{xxx}= & 0 \; .
\end{split}
\end{equation}
One defines here $u=\sqrt{2}\, w$, $h=v'$ and changes the spatial
variable to $X=x/\sqrt{2}$. This casts the
approximate system \eqref{appro-sys2}
into the canonical Kaup-Boussinesq form \cite{Kau76}
\begin{equation}\label{KB-canonical}
\begin{split}
u_t+u u_X + h_X = & 0 \; ,\\
h_t+(h u)_X-\tfrac14 u_{XXX} = & 0 \; .
\end{split}
\end{equation}
Again a similar approximation can be derived when $v(x,t)$ is close to
$-1$ and $w(x,t)$ small.

\section{Periodic solutions and Whitham equations}\label{sec3}

Among the key elements that are generated during the evolution of
nonlinear waves, an important role is played by the DSWs that can be
represented as modulated periodic solutions of the corresponding
nonlinear wave equation.  Consequently, for
classifying of the wave patterns evolving from an initial
discontinuity in the polarization mode, we have to present the
periodic solutions of the LL equation in the most convenient
form and to derive the corresponding Whitham modulation equations.

In Ref.~\cite{Con16} the periodic solutions have been found by a
direct method, and were not parameterized in terms of Riemann
invariants. The Whitham equations were used in El's form \cite{el-05}
which provides the main informations concerning the evolution of
initial step-like discontinuous distributions without requiring the
knowledge of the Riemann invariants. However, for solving the full
Riemann problem it is more appropriate to use methods based on the
explicit knowledge of the Riemann invariants. In this section we shall
obtain the periodic solutions of the LL equation by means of the
finite gap integration method, give the explicit form of the Riemann
invariants, and derive the corresponding Whitham modulation equations.

\subsection{One-phase finite-gap integration method of the easy-plane
  Landau-Lifshitz equation}

As well known, the LL equation \eqref{syst2} is integrable by the inverse
scattering transform method (see, e.g., \cite{Bor78,Bor88}). The
corresponding Lax pair can be written as
\begin{align}
\label{lax1}
  &\frac{\partial}{\partial x}
\begin{pmatrix}
\psi_1 \\
\psi_2 \\
\end{pmatrix}
=\left(
   \begin{array}{cc}
     F & G \\
     H & -F \\
   \end{array}
 \right)
\begin{pmatrix}
\psi_1 \\
\psi_2 \\
\end{pmatrix}
\;, \\
\label{lax2}
&\frac{\partial}{\partial t}
\begin{pmatrix}
\psi_1 \\
\psi_2 \\
\end{pmatrix}
=\left(
   \begin{array}{cc}
     A & B \\
     C & -A \\
   \end{array}
 \right)
\begin{pmatrix}
\psi_1 \\
\psi_2 \\
\end{pmatrix}
 \;,
\end{align}
where
\begin{equation}\label{lax-u}
\begin{split}
  F&=\frac{{\rm i}\lambda}2M_z,\quad G=-\frac{1}2\sqrt{1-\lambda^2}M_-,\\
H& =-\frac{1}2\sqrt{1-\lambda^2}M_+,\\
  A&=\frac{\rm i}2(1-\lambda^2)M_z+\frac{\lambda}4[(M_-)_x
M_+-M_-(M_+)_x],\\
  B&=\frac{1}2\lambda\sqrt{1-\lambda^2}M_-+\\
& \frac{\rm i}2\sqrt{1-\lambda^2}[(M_z)_x
M_- -M_z(M_-)_x],\\
  C&=\frac{1}2\lambda\sqrt{1-\lambda^2}M_+
-\\
& \frac{\rm i}2\sqrt{1-\lambda^2}[(M_z)_x M_+ -M_z(M_+)_x].
  \end{split}
\end{equation}
In \eqref{lax-u} $M_{\pm}=M_x\pm {\rm i}M_y$ and $\lambda$ is a constant
spectral parameter. Periodic solutions for ferromagnets with an
easy-axis anisotropy were found in Ref. \cite{Kam92} and we shall here
adapt the approach used in this reference
to the easy-plane case of Eq.~\eqref{syst2}.  The
$2\times2$ linear problems \eqref{lax1} and
\eqref{lax2} have two linearly independent
basis solutions which we denote as $(\psi_1,\,\psi_2)^T$ and
$(\vphi_1,\,\vphi_2)^T$. We define the ``squared basis functions''
\begin{equation}\label{basis}
  f=-\frac{\rm i}2(\psi_1\varphi_2+\psi_2\varphi_1), \quad
g=\psi_1\varphi_1,\quad h=-\psi_2\varphi_2,
\end{equation}
which obey the linear equations
\begin{subequations}\label{fzeta_x}
\begin{align}
f_{x}=& -{\rm i}H g+{\rm i}G h \;,\label{fzxa} \\
g_x=& 2{\rm i}G f+2F g \;,\label{fzxb} \\
h_x= &-2{\rm i}H f-2F h \;,\label{fzxc}
\end{align}
\end{subequations}
and
\begin{subequations}\label{fzeta_t}
\begin{align}
f_t=& -{\rm i}C g+{\rm i}B h \;,\label{fzta} \\
g_t=& 2{\rm i}B f+2A g \;,\label{fztb} \\
h_t=&-2{\rm i}C f-2A h \;.\label{fztc}
\end{align}
\end{subequations}

It is easy to check that the expression $f^2-gh$ does not depend on
$x$ and $t$, however it can depend on the spectral parameter $\la$.
The (quasi)periodic solutions are distinguished by the condition that
the term $f^2-gh$ be a polynomial $P(\lambda)$.  For the one-phase
case which we are interested in, it suffices to consider a fourth degree
polynomial
\begin{equation}\label{pol-P}
\begin{split}
  f^2-gh=P(\la)= & \prod_{i=1}^4(\la-\la_i)\\
= & \la^4-s_1\la^3+s_2\la^2-s_3\la+s_4,
\end{split}
\end{equation}
where $s_i$ are standard symmetric functions of the four zeros
($\la_1$, $\la_2$, $\la_3$ and $\la_4$) of the polynomial:
\begin{equation}\label{symm-func}
\begin{split}
  &s_1=\sum_i\la_i,\quad s_2=\sum_{i<j}\la_i\la_j,
  \quad s_3=\sum_{i<j<k}\la_i\la_j\la_k, \\
  & s_4=\la_1\la_2\la_3\la_4.
  \end{split}
\end{equation}
We write the solution of Eqs.~(\ref{fzeta_x}) and \eqref{fzeta_t}
under the form
\begin{equation}\label{sol:fgh}
\begin{split}
&f(x,t)=M_z \la^2-f_1(x,t)\la+f_2(x,t) \;,\\
&g(x,t)=M_-\sqrt{1-\la^2}\left(\lambda-\mu(x,t)\right) \;,\\
&h(x,t)=M_+\sqrt{1-\la^2}\left(\lambda-\mu^*(x,t)\right) \;,
\end{split}
\end{equation}
where, to simplify computations, we have chosen the coefficients of
the terms with the highest degrees in $\la$ in such a way that the
identity (\ref{pol-P}) is already satisfied at order $\la^4$. The
quantity $f_1(x,t)$, $f_2(x,t)$, $\mu(x,t)$ and $\mu^*(x,t)$ in
\eqref{sol:fgh} are yet unknown functions;  $\mu(x,t)$ and
$\mu^*(x,t)$ are {\it a priori} unrelated, but we
shall soon establish that they are complex conjugate one to the other,
whence the notation.

Plugging expressions \eqref{sol:fgh} back
into \eqref{pol-P} and equating the coefficients of the powers of
$\la$ yields four conservation laws
\begin{equation}\label{268.6}
\begin{split}
&-2 f_1 w+(1-w^2)(\mu+\mu^*)=s_1 \,,\\
&2f_1f_2-(1-w^2)(\mu+\mu^*)=s_3 \,,\\
&f_1^2-2 f_2w+(1-w^2)(\mu\,\mu^*-1)=s_2 \,,\\
&f_2^2-(1-w^2)\, \mu\,\mu^*=s_4 \,,
\end{split}
\end{equation}
where we have used the above defined notation $w\equiv -M_z$ and have
also taken into account the normalization
\begin{equation}\label{normal}
M_+M_-+M_z^2=1\; .
\end{equation}
Substitution of (\ref{sol:fgh}) into (\ref{fzeta_x}) and
\eqref{fzeta_t} gives, after equating the coefficients of powers of
$\la$, a number of differential equations; we shall write down here
the ones which are the most important for our purpose. For instance
the equation \eqref{fzxa} gives
\begin{equation}\label{269.8}
  w_x=-\frac{\rm i}{2}(1-w^2)(\mu-\mu^*),\quad f_{1,x}=0,\quad f_{2,x}=w_x.
\end{equation}
 After factoring out the term
$\sqrt{1-\la^2}$, the equality of the coefficients of the terms of
order $\la$ in both sides of equation \eqref{fzxb} yields
\begin{equation}\label{269.10}
  (M_-)_x/M_-={\rm i}(f_1+w\mu).
\end{equation}
This equation, with account of $\phi_x=v$, $M_-=-\sqrt{1-w^2}\exp(-{\rm i}
\phi)$  and of the first of Eqs.~\eqref{269.8}, leads to
the following expression for the relative velocity:
\begin{equation}\label{270.14}
  v=-f_1-\frac12(\mu+\mu^*)w.
\end{equation}

The variable $\mu$ satisfies the equation
\begin{equation}\label{269.7}
  \mu_x={\rm i}\sqrt{P(\mu)}
\end{equation}
which can be easily obtained by putting the free parameter $\la$ equal
to $\mu$ in equation \eqref{fzxb}.
Substitution of (\ref{269.10}) and (\ref{269.8}) into \eqref{fztb}
where the parameter $\la$ is taken equal to
$\mu$ gives, owing to the first of identities (\ref{268.6}),
the equation $\mu_t=-({\rm i}/2)s_1\sqrt{P(\mu)}=-(1/2)s_1\mu_x$. This
indicates that $\mu$ depends on the variable $\xi=x-(s_1/2)t$
only, that is
\begin{equation}\label{269.13}
  \mu_{\xi}={\rm i}\sqrt{P(\mu)}\; ,\quad \xi=x-Vt\; ,\quad V=\frac12 s_1.
\end{equation}
Formally, equation (\ref{269.13}) can be solved in terms of elliptic
functions and it is then parameterized by the zeroes of the polynomial
$P(\la)$. However, even for given values of these zeroes, the
trajectory of $\mu$ in the complex $\mu$-plane is not known and
therefore it is impossible to prescribe the initial value of $\mu$
without some additional study. This difficulty can be overcome by the
method suggested in Ref. \cite{kamch-90}, according to which the parameters
$f_1,\,f_2,\,\mu,\,\mu^*$ are to be represented as functions of $w$.
This yields the solution in a so-called ``effective'' form, not
subject to any additional constraint.

After simple manipulations on the system (\ref{268.6}), we find, for a
given set of $\la_i$ ($i=1,2,3,4$), four possible forms of $f_1$:
\begin{subequations}\label{270.16}
\begin{equation}\label{270.16a}
  f_1=\pm\sqrt{(1+s_2+s_4+s_4')/2},
\end{equation}
and
\begin{equation}\label{270.16b}
  f_1=\pm\mathrm{sgn}(s_1+s_3)\sqrt{(1+s_2+s_4-s_4')/2},
\end{equation}
\end{subequations}
where we have defined
\begin{equation}\label{270.15}
  \la_i'=\sqrt{1-\la_i^2}\; ,\quad  s_4'=\la_1'\la_2'\la_3'\la_4',
\end{equation}
and made use of the identity $(1+s_2+s_4)^2-(s_1+s_3)^2=(s_4')^2$.  The
factor $\mathrm{sgn}(s_1+s_3)$ is introduced for making $f_1$  (and its
derivatives with respect to $\la_i$)  continuous functions of
$\la_i$. For $f_2$ we obtain in all cases
\begin{equation}\label{270.17}
  f_2=(s_1+s_3)/2f_1+w,
\end{equation}
and the variables $\mu,\,\mu^*$ are given by the expressions
\begin{equation}\label{270.18}
  \mu,\mu^*=\frac{s_1+2f_1w\pm2\, {\rm i}\sqrt{-\mathcal{R}(w)}}{2(1-w^2)},
\end{equation}
where
\begin{equation}\label{sol:muS}
\begin{split}
\mathcal{R}(w) = &w^4+ \frac{s_1+s_3}{f_1} w^3 + s_2 w^2+ \left(f_1 s_1
-\frac{s_1+s_3}{f_1} \right) w\\
& + \frac{1}{4} \left(s_1^2-4-4s_2+4f_1^2\right).
\end{split}
\end{equation}
Since $\mu$ depends on $\xi$ only, the same holds for $w$, which, as
follows from Eqs.~(\ref{269.8}) and (\ref{270.18}), satisfies the
equation
\begin{equation}\label{271.20}
  w_{\xi}=\sqrt{-\mathcal{R}(w)}
\end{equation}
This equation admits a real solution when $w$ oscillates between two
of the zeroes of $\mathcal{R}(w)$ (provided they both are located in
the interval $[-1,1]$), in a domain where $\mathcal{R}(w)\leq 0$, and
in this case, on sees from \eqref{270.18} that $\mu$ and $\mu^*$ are
complex conjugated variables, as was anticipated earlier.

Actually, Eq.~(\ref{271.20}) coincides with Eq. (30) of Ref.
\cite{{Con16}} (with $Q(w)$ replaced by $\mathcal{R}(w)$) and we shall
reproduce here briefly its solutions for convenience and future
references.  We denote the zeroes of $\mathcal{R}$ as $w_1\leq w_2\leq
w_3 \leq w_4$.

(A) We first consider the periodic solution corresponding to oscillations
of $w$ in the interval
\begin{equation}\label{eq18}
w_1\leq w\leq w_2.
\end{equation}
In this case the solution of Eq.~(\ref{271.20}) can be written as
\begin{equation}\label{eq20}
w=w_2-\frac{(w_2-w_1)\cn^2(W,m)}{1+\frac{w_2-w_1}{w_4-w_2}\sn^2(W,m)},
\end{equation}
where it is assumed that $w(0)=w_1$,
\begin{equation}\label{eq21}
W=\sqrt{(w_3-w_1)(w_4-w_2)}\,\xi/2,
\end{equation}
and
\begin{equation}\label{eq22}
m=\frac{(w_4-w_3)(w_2-w_1)}{(w_4-w_2)(w_3-w_1)},
\end{equation}
$\cn$ and $\sn$ being Jacobi elliptic functions \cite{Abram}.
The wavelength of the oscillating function \eqref{eq20} is
\begin{equation}\label{eq23}
L=\frac{4K(m)}{\sqrt{(w_3-w_1)(w_4-w_2)}},
\end{equation}
where $K(m)$ is the complete elliptic integral of the first kind
\cite{Abram}.  In the limit $w_3\to w_2$ ($m\to1$) the wavelength
tends to infinity and the solution (\ref{eq20}) transforms to a
soliton
\begin{equation}\label{eq24}
w=w_2-\frac{w_2-w_1}{\cosh^2W+\frac{w_2-w_1}{w_4-w_2}\sinh^2W}.
\end{equation}
This is a ``dark soliton'' for the variable $w$.

The limit $m\to0$ can be reached in two ways.

(i) If $w_2\to w_1$, then we get
\begin{equation}\label{eq25}
\begin{split}
w&\cong w_2-\frac12(w_2-w_1)\cos[k(x-V t)],\\
k&=\sqrt{(w_3-w_1)(w_4-w_1)}.
\end{split}
\end{equation}
This is a small-amplitude limit describing propagation of a harmonic wave.

(ii) If $w_4=w_3$ but $w_1\neq w_2$, then we get a nonlinear wave
represented in terms of trigonometric functions:
\begin{equation}\label{eq26}
\begin{split}
 w&=w_2-\frac{(w_2-w_1)\cos^2W}{1+\frac{w_2-w_1}{w_3-w_2}\sin^2W},\\
 W&=\sqrt{(w_3-w_1)(w_3-w_2)}\,\xi/2.
\end{split}
\end{equation}
If we take the limit $w_2-w_1\ll w_3-w_1$ in this solution, then we
return to the small-amplitude limit (\ref{eq25}) with $w_4=w_3$. On
the other hand, if we take here the limit $w_2\to w_3=w_4$, then the
argument of the trigonometric functions becomes small and we can
approximate them by the first terms of their series expansions. This
corresponds to an algebraic soliton of the form
\begin{equation}\label{eq27}
w=w_2-\frac{w_2-w_1}{1+(w_2-w_1)^2(x-Vt)^2/4}.
\end{equation}

(B) In the second case, the variable $w$ oscillates in the interval
\begin{equation}\label{eq28}
w_3\leq w\leq w_4\; .
\end{equation}
Here again, a standard calculation yields
\begin{equation}\label{eq30}
w=w_3+\frac{(w_4-w_3)\cn^2(W,m)}{1+\frac{w_4-w_3}{w_3-w_1}\sn^2(W,m)}.
\end{equation}
with the same definitions (\ref{eq21}), (\ref{eq22}), and (\ref{eq23})
for $W$, $m$, and $L$, respectively, and $w(0)=w_4$.
In the soliton limit $w_3\to w_2$ ($m\to1$) we get
\begin{equation}\label{eq31}
w=w_2+\frac{w_4-w_2}{\cosh^2W+\frac{w_4-w_2}{w_2-w_1}\sinh^2W}.
\end{equation}
This is a ``bright soliton'' for the variable $w$.

Again, the limit $m\to0$ can be reached in two ways.

(i) If $w_4\to w_3$, then we obtain a small-amplitude harmonic wave
\begin{equation}\label{eq32}
\begin{split}
w&\cong w_3+\frac12(w_4-w_3)\cos[k(x-V t)],\\
k&=\sqrt{(w_3-w_1)(w_3-w_1)}.
\end{split}
\end{equation}
This is a small-amplitude limit describing a harmonic wave.

(ii) If $w_2=w_1$, then we obtain another nonlinear trigonometric solution,
\begin{equation}\label{eq33}
\begin{split}
w&=w_3+\frac{(w_4-w_3)\cos^2W}{1+\frac{w_4-w_3}{w_3-w_1}\sin^2W},\\
 W&=\sqrt{(w_3-w_1)(w_4-w_1)}\,\xi/2.
\end{split}
\end{equation}
If we assume that $w_4-w_3\ll w_4-w_1$, then this reproduces
the small-amplitude limit (\ref{eq32}) with $w_2=w_1$. On the other hand,
in the limit $w_3\to w_2=w_1$ we obtain the algebraic soliton solution:
\begin{equation}\label{eq34}
w=w_1+\frac{w_4-w_1}{1+(w_4-w_1)^2(x-Vt)^2/4}.
\end{equation}

The solutions presented above are parameterized by the zeroes $w_i$
($i=1,2,3,4$) of the polynomial (\ref{sol:muS}) whose coefficients are
expressed in terms of the zeroes $\la_i$ of the polynomial $P(\la)$ which
plays a key role in the finite-gap integration method. As we shall
see, the parameters $\la_i$ represent the Riemann invariants in the
Whitham modulation theory.  We want to express the solutions of
\eqref{271.20} in terms of these Riemann invariants: we therefore need
to express the $w_i$'s in terms of the $\la_i$'s explicitly, without
having to solve the algebraic equation $\mathcal{R}(w)=0$. This has
been already achieved in Ref.~\cite{Kam92}, but we shall derive here
expressions under a form which is more convenient for
subsequent applications. To this end, we rewrite the identity
(\ref{pol-P}) using explicit formulas for the functions $f$, $g$ and $h$:
\begin{equation}\nonumber
\begin{split}
  &P(\la)=\left(-w\la^2-f_1\la+\frac{s_1+s_3}{2f_1}+w\right)^2\\
& -(1-w^2)(1-\la^2) \times
\left[\la-\frac{s_1+2f_1w+2\sqrt{\mathcal{R}(w)}}{2(1-w^2)}\right]\\
&\times
\left[\la-\frac{s_1+2f_1w-2\sqrt{\mathcal{R}(w)}}{2(1-w^2)}\right].
  \end{split}
\end{equation}
Let $\la=\la_i$ be a zero of the polynomial $P(\la)$ and $w$ be also
one of the zeros of $\mathcal{R}(w)$. Then, for a given $w$, the above
identity yields two equations
\begin{equation}\nonumber
  \begin{split}
&  -w\la_i^2-f_1\la_i+\frac{s_1+s_3}{2f_1}+w
  =\\
& \pm\sqrt{1-w^2}\la_i'\left[\la_i-\frac{s_1+2f_1w}{2(1-w^2)}\right]
  \end{split}
\end{equation}
for the four roots $\la_i$.  We assume that $\la_1$ and $\la_2$
correspond to the upper sign and that $\la_3$ and $\la_4$ correspond
to the lower sign. We temporarily introduce the notation
\begin{equation}\label{272.28}
  \n{i}=2f_1w\la_i'+(s_1+s_3-2f_1^2\la_i)/\la_i'=2f_1\la_i'w+\tilde{s}_i,
\end{equation}
where $\tilde{s}_i$ is a notation for
$(s_1+s_3-2f_1^2\la_i)/\la_i'$. The use of formulas (\ref{270.16}) for
$f_1$ yields
\begin{equation}\label{272.27}
  \ts_i=(s_1-\la_i)\la_i'+s_4\frac{\la_i'}{\la_i}\mp s_4'\frac{\la_i}{\la_i'},
\end{equation}
where the upper sign corresponds to Eq.~(\ref{270.16a}), and the lower
one to Eq.~(\ref{270.16b}. It is then possible to rewrite
Eq.~\eqref{272.28} under the form
\begin{equation}\label{271.24}
  \n{i}=\pm\frac{f_1}{\sqrt{1-w^2}}\left[2\la_i(1-w^2)-s_1-2f_1w\right].
\end{equation}
Dividing expressions \eqref{271.24} for the $\n{i}$'s
one by the other for various pairs of $i$ and $j\, (\ne i)$, we get
\begin{equation}\nonumber
  \frac{\n{i}}{\n{j}}=
\pm\frac{2\la_i(1-w^2)-s_1-2f_1w}{2\la_j(1-w^2)-s_1-2f_1w},
\end{equation}
where the plus signs applies for $\n{1}/\n{2}$ and $\n{3}/\n{4}$ and
the minus one for the other choices of these pairs. Consequently we
have
\begin{equation}\label{272.26}
  1-w^2=\frac{s_1+2f_1w}{2}\cdot
\frac{\n{i}\pm\n{j}}{\la_j\n{i}\pm\la_i\n{j}}
\end{equation}
with the same sign convention. Equating these expressions for
$1-w^2$ to each other, we obtain a number of equations --- linear and
quadratic in the $\n{i}$'s.  For example, from the equality
\begin{equation}\label{noname}
  \frac{\n{1}+\n{2}}{\la_2\n{1}+\la_1\n{2}}=
\frac{\n{2}-\n{3}}{\la_3\n{2}-\la_2\n{3}}
\end{equation}
we get the first relationship of the system
\begin{equation}\label{272.29}
  \begin{split}
  &(\la_3-\la_2)\n{1}+(\la_3-\la_1)\n{2}-(\la_2-\la_1)\n{3}=0,\\
  &(\la_3-\la_2)\n{1}+(\la_3-\la_1)\n{2}+(\la_2-\la_1)\n{3}=0,\\
  &(\la_3-\la_2)\n{1}-(\la_3-\la_1)\n{2}+(\la_2-\la_1)\n{3}=0,\\
  &(\la_3-\la_2)\n{1}-(\la_3-\la_1)\n{2}-(\la_2-\la_1)\n{3}=0,\\
  \end{split}
\end{equation}
and the three others can be obtained by considering equalities of the type
\eqref{noname} for other choices of pairs of indices.  Although these
equations are not linearly independent (one can check that one of them
is a linear combination of the three others), we prefer to deal with
all of them to get symmetrical expressions for all four roots of the
resolvent. Indeed, using the expression (\ref{272.28}) for $\n{i}$,
each of the relationships \eqref{272.29} becomes a linear equation for
$w$ and yields one of the zeroes of the polynomial $\mathcal{R}(w)$.
As a result we obtain the formulae
\begin{equation}\label{272.30}
  \begin{split}
  &w_1=-\frac1{2f_1}\cdot
\frac{(\la_3-\la_2)\ts_1+(\la_3-\la_1)\ts_2-(\la_2-\la_1)\ts_3}
  {(\la_3-\la_2)\la_1'+(\la_3-\la_1)\la_2'-(\la_2-\la_1)\la_3'},\\
  &w_2=-\frac1{2f_1}\cdot
\frac{(\la_3-\la_2)\ts_1+(\la_3-\la_1)\ts_2+(\la_2-\la_1)\ts_3}
  {(\la_3-\la_2)\la_1'+(\la_3-\la_1)\la_2'+(\la_2+\la_1)\la_3'},\\
  &w_3=-\frac1{2f_1}\cdot
\frac{(\la_3-\la_2)\ts_1-(\la_3-\la_1)\ts_2-(\la_2-\la_1)\ts_3}
  {(\la_3-\la_2)\la_1'-(\la_3-\la_1)\la_2'-(\la_2-\la_1)\la_3'},\\
  &w_4=-\frac1{2f_1}\cdot
\frac{(\la_3-\la_2)\ts_1-(\la_3-\la_1)\ts_2+(\la_2-\la_1)\ts_3}
  {(\la_3-\la_2)\la_1'-(\la_3-\la_1)\la_2'+(\la_2-\la_1)\la_3'},\\
  \end{split}
\end{equation}
where the $w_i$'s are ordered according to $w_1\leq w_2\leq w_3\leq
w_4$ under the suppositions that $\la_1\leq\la_2 \leq\la_3 \leq\la_4$
and $f_1>0$. A change of sign of $f_1$ leads to a simple reordering of
the expressions for the $w_i$'s. We see that the zeroes $\la_i$ of the
polynomial $P(\la)$ and the zeroes $w_i$ of the polynomial
$\mathcal{R}(w)$ are related by the symmetrical formulae
(\ref{272.30}), therefore we shall call $\mathcal{R}(w)$ the resolvent
of the polynomial $P(\la)$ (for example, in the case of NLS equation
an analogous method yields the well-known Ferrari cubic resolvent used
for solving in radicals fourth degree algebraic
equations). Formulae \eqref{272.30} are equivalent to those obtained
in Ref.~\cite{Kam92}, however they are more convenient for the study
of degenerate cases presented below. It is important to note that we
have four values of $f_1$ for each set of the $\la_i$'s (and corresponding
values (\ref{272.27}) for $\ts_i$), which are
thus mapped by the formulae (\ref{272.30}) to four sets of
$w_i$'s. This multiplicity of mappings will prove of tremendous importance
when applying the Whitham theory of modulations to the representations
\eqref{eq20} and \eqref{eq30} of the periodic solutions. Before addressing this
crucial question we first
need to demonstrate that the parameters $\la_i$ ($i=1,2,3,4$) are the
Riemann invariants of the Whitham system for the averaged conservation
laws. This is achieved in the next section.

\subsection{Whitham equations}

In modulated waves the $\la_i$'s become slowly varying functions of
the space and time variables and their evolution is governed by the
Whitham modulation equations. Whitham showed in
Refs.~\cite{whitham-65,Whithbook} that these equations can be obtained
by averaging the conservation laws of the full nonlinear system over
fast oscillations (whose wavelength $L$ changes slowly along the total
wave pattern).  Generally speaking, in cases where the periodic
solution is characterized by four parameters, this averaging procedure
leads to a system of four equations of the type
$w_{i,t}+\sum_jv_{ij}(w_1,w_2,w_3,w_4)w_{j,x}=0$ with 16 entries of
the ``velocity matrix'' $v_{ij}$.  However, the Landau-Lifshitz
equation being completely integrable, this system of four equations
reduces to a diagonal ``Riemann form'' for the $\la_i$'s, similarly to
what occurs for the usual Riemann invariants of non-dispersive waves
(see, e.g., Ref. \cite{LL-6}). As a result, the $\la_i$'s are called the Riemann
invariants of the dispersive nonlinear wave.  We shall study their
properties by using the method devised in Refs.~\cite{kamch-94,Kam00}.

First of all, we notice that Eq.~(\ref{270.18}) implies that, during
the oscillations of $w$, the variable $\mu$ describes a cycle in the
complex plane which encloses either points $\la_1$ and $\la_2$ or
points $\la_3$ and $\la_4$ (according to Eq.~(\ref{269.7}) the
variable $\mu$ runs along one of the two loops of an hyperelliptic
curve while the $w$-variable oscillates within the corresponding
interval).  Hence, from Eq.~(\ref{269.13}) one can derive the
following expression for the wavelength
\begin{equation}\label{276.49}
  L=\oint\frac{d\mu}{\sqrt{-P(\mu)}}
=\frac{4K(m)}{\sqrt{(\la_3-\la_1)(\la_4-\la_2)}},
\end{equation}
Comparison of this expression with Eq.~(\ref{eq23}) leads to the identities
\begin{equation}\label{276.50a}
m=\frac{(w_4-w_3)(w_2-w_1)}{(w_4-w_2)(w_3-w_1)}
  =\frac{(\la_4-\la_3)(\la_2-\la_1)}{(\la_4-\la_2)(\la_3-\la_1)},
\end{equation}
and
\begin{equation}\label{276.50b}
(w_4-w_2)(w_3-w_1)=(\la_4-\la_2)(\la_3-\la_1).
\end{equation}

From \eqref{fzxb} and \eqref{fztb} and owing to the normalization
condition \eqref{normal} one gets $B g_x-G g_t=2(B F -A G)g= \tfrac12
\sqrt{1-\la^2} [{\rm i}\, M_z M_-+\la (M_-)_x]g$. Using the equations
of motion \eqref{syst2} this last term can be rewritten as $g G_t-g
B_x$. Dividing by $g^2$, one can cast the resulting identity under the
form
\begin{equation}\label{275.47}
  \frac{\prt}{\prt t}\left(\sqrt{P(\la)}\cdot\frac{G(\la)}{g(\la)}\right)-
  \frac{\prt}{\prt x}\left(\sqrt{P(\la)}\cdot\frac{B(\la)}{g(\la)}\right)
=0.
\end{equation}
We shall use this equation as the generating function of the
conservation laws of the Landau-Lifshitz equation: a series expansion
in inverse powers of $\la$ gives an infinite number of conservation
laws of this completely integrable system. The factor $\sqrt{P(\la)}$
has been introduced to transform the identity (\ref{pol-P}) to the
form
\begin{equation}\nonumber
  \left(\frac{f}{\sqrt{P(\la)}}\right)^2-
\frac{g}{\sqrt{P(\la)}}\cdot\frac{h}{\sqrt{P(\la)}}=1,
\end{equation}
so that the right-hand side is independent of the variations of
$\la_i$ in a modulated wave, hence the densities and fluxes in the
conservation laws can change due to modulations only, as it should be,
and any changes caused by $\lambda$-dependent normalization of the
$f,g,h$-functions are excluded.

Substitution of Eqs.~(\ref{lax-u}) and (\ref{sol:fgh})
into (\ref{275.47}) and its simple transformation with the use of
Eqs.~(\ref{268.6}) and (\ref{269.10}) gives
\begin{equation}\nonumber
  \frac{\prt}{\prt t}\left(\frac{\sqrt{P(\la)}}{\la-\mu}\right)+
  \frac{\prt}{\prt x}
\left[\sqrt{P(\la)}\left(1+\frac{s_1/2}{\la-\mu}\right)\right]=0,
\end{equation}
Averaging of the density and of the flux in this expression over one
wavelength $L$ (\ref{276.49}) yields the generating function of the
{\it averaged} conservation laws:
\begin{equation}\label{276.48}
\begin{split}
  &\frac{\prt}{\prt t}
\left[\frac{\sqrt{P(\la)}}L\oint\frac{d\mu}{(\la-\mu)\sqrt{-P(\mu)}}\right]\\
  &+
  \frac{\prt}{\prt x}
\left[\frac{\sqrt{P(\la)}}L\oint\left(1+\frac{s_1/2}{\la-\mu}\right)
  \frac{d\mu}{\sqrt{-P(\mu)}}\right]=0.
  \end{split}
\end{equation}
The condition that in the limit $\la\to\la_i$ the singular terms
cancel yields
\begin{equation}\label{equa65}
\begin{split}
&  \oint\frac{d\mu}{(\la_i-\mu)\sqrt{-P(\mu)}}\cdot\frac{\prt\la_i}{\prt t}
  +\\
& \left(L+\frac{s_1}2\oint\frac{d\mu}{(\la_i-\mu)\sqrt{-P(\mu)}}\right)
\cdot\frac{\prt\la_i}{\prt x}=0.
\end{split}
\end{equation}
From the definition \eqref{276.49} of $L$ one obtains
$$
\oint \frac{d\mu}{(\la_i-\mu)\sqrt{-P(\mu)}}=-2\, \frac{\prt L}{\prt\la_i},
$$
which makes it possible to cast Eq.~\eqref{equa65} under the form of a
Whitham equation for the variables $\la_i$:
\begin{equation}\label{276.56}
  \frac{\prt\la_i}{\prt t}+v_i\frac{\prt\la_i}{\prt x}=0,
\end{equation}
where the Whitham velocity $v_i$ is given by
\begin{equation}\label{276.57}
  v_i=\frac{s_1}2-\frac{L}{2\prt L/\prt\la_i},\quad \mbox{for}\quad
i\in\{1,2,3,4\}.
\end{equation}
By means of Eq.~(\ref{276.49}) one obtains the following
explicit expressions
\begin{equation}\label{276.55}
\begin{split}
    & v_1=\frac{1}{2}\sum_{i=1}^{4}\lambda_i
-\frac{(\lambda_4-\lambda_1)(\lambda_2-\lambda_1)
K(m)}{(\lambda_4-\lambda_1)K(m) -(\lambda_4-\lambda_2)E(m)}, \\
    & v_2=\frac{1}{2}\sum_{i=1}^{4}\lambda_i
+\frac{(\lambda_3-\lambda_2)(\lambda_2-\lambda_1)K(m)}
{(\lambda_3-\lambda_2) K(m)-(\lambda_3-\lambda_1)E(m)}, \\
    & v_3=\frac{1}{2}\sum_{i=1}^{4}\lambda_i-
\frac{(\lambda_4-\lambda_3)(\lambda_3-\lambda_2)K(m)}
{(\lambda_3-\lambda_2)K(m)- (\lambda_4-\lambda_2)E(m)}, \\
    & v_4=\frac{1}{2}\sum_{i=1}^{4}\lambda_i
+\frac{(\lambda_4-\lambda_2)(\lambda_4-\lambda_1)K(m)}
{(\lambda_4-\lambda_1)K(m) -(\lambda_3-\lambda_1)E(m)},
    \end{split}
\end{equation}
where $K(m)$ and $E(m)$ are complete elliptic integrals of the first
and second kind, respectively.

\begin{figure}
  \centering
\includegraphics[width=\linewidth]{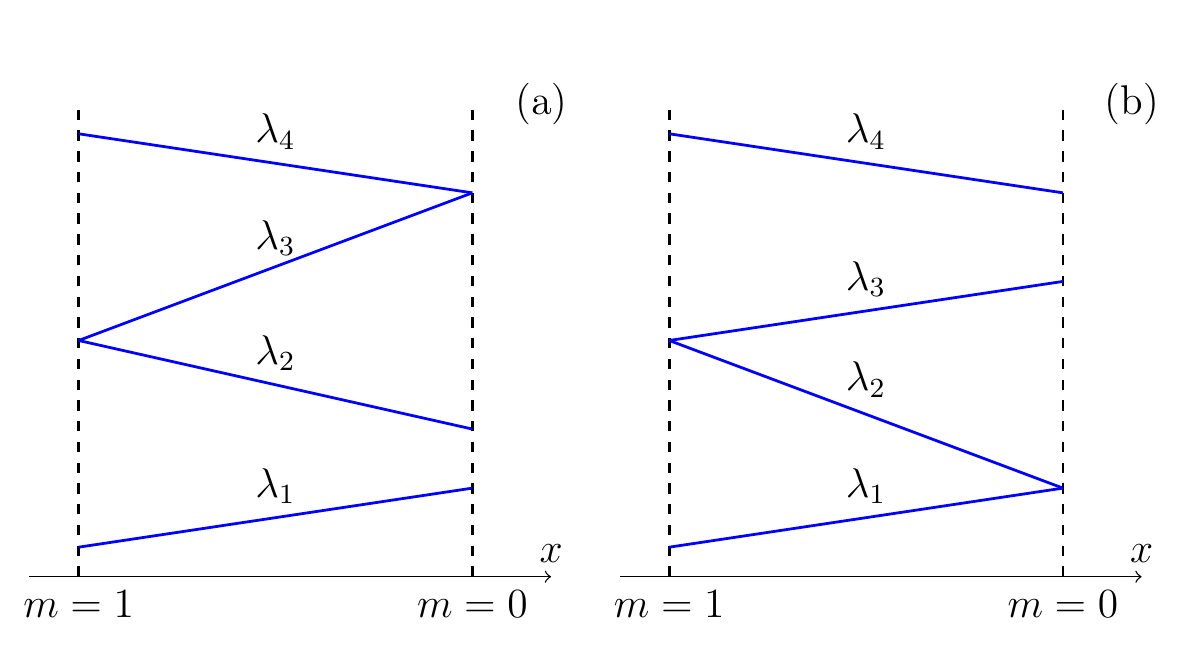}
  \caption{Sketches of the space dependence of the Riemann invariants
    along a DSW. In both cases the limit $\la_2=\la_3$ corresponds to
    the soliton edge. The polarity of the solitons depends on the
    choice of solution of formulae (\ref{272.30}) by which the
    solution of the Whitham equations are mapped onto the parameters
    $w_i$.  The small amplitude edge corresponds to $\la_3=\la_4$ in
    case (a) and to $\la_1=\la_2$ in case (b).}
    \label{fig1}
\end{figure}
In a modulated wave representing a DSW, the Riemann invariants change
with $x$ and $t$.  The DSW occupies a region in space at the edges of
which two Riemann invariants coincide. There are two possible
situations represented schematically in Fig.~\ref{fig1}. In both cases
the soliton edge corresponds to $\la_3=\la_2$ $(m=1)$ and at this edge
the Whitham velocities are given by
\begin{equation}\label{277.56}
\begin{split}
  &v_1=\frac12(3\la_1+\la_4),\quad v_2=v_3=\frac12(\la_1+2\la_2+\la_4),\\
  &v_4=\frac12(\la_1+3\la_4),\quad\text{for}\quad \la_3=\la_2.
  \end{split}
\end{equation}
The small amplitude limit $m=0$ can be obtained in two ways. If
$\la_3=\la_4$ (Fig.~\ref{fig1}(a)), then we get
\begin{equation}\label{277.57}
\begin{split}
  &v_1=\frac12(3\la_1+\la_2),\quad v_2=\frac12(\la_1+3\la_2),\\
  &v_3=v_4=2\la_4+\frac{(\la_2-\la_1)^2}{2(\la_1+\la_2-2\la_4)},
  \end{split}
\end{equation}
and if $\la_2=\la_1$ (Fig.~\ref{fig1}(b)), then
\begin{equation}\label{277.58}
\begin{split}
  &v_1=v_2=2\la_1+\frac{(\la_4-\la_3)^2}{2(\la_3+\la_4-2\la_1)},\\
  &v_3=\frac12(3\la_3+\la_4),\quad v_4=\frac12(\la_3+3\la_4).
  \end{split}
\end{equation}

As one can see from Eqs.~(\ref{272.30}), for $\la_2=\la_1$ we have
$w_2=w_1$ and for $\la_3=\la_4$ we have $w_3=w_4$. Consequently,
Eqs.~(\ref{277.57}) and (\ref{277.58}) represent also the Whitham
velocities for evolution of shocks approximated by the trigonometric
solutions (\ref{eq33}) and (\ref{eq26}), respectively. We shall
call them ``trigonometric shocks''. As we shall see, they play an
important role in the classification of the possible wave structures
evolving from initial discontinuities.

We can now proceed to the description of key elements (``building blocks'')
from which the wave patterns are constructed.

\section{Key elements}\label{sec4}

The Riemann problem we consider in the present work consists in the
study of the time evolution of an initial step-like structure of the
form
\begin{equation}\label{init-cond}
\begin{split}
w(x,t=0)=\begin{cases}
w_L & \mbox{if}\;\; x<0 \; , \\
w_R & \mbox{if}\;\; x>0\; ,
\end{cases}
\\
v(x,t=0)=\begin{cases}
v_L & \mbox{if}\;\; x<0 \; ,\\
v_R & \mbox{if}\;\; x>0\; .
\end{cases}
\end{split}
\end{equation}
We consider the hyperbolic case where the four boundary values $w_L$,
$w_R$, $v_L$ and $v_R$ are contained in the interval $[-1,1]$.  The
initial distribution involves no characteristic constants having the
dimension of a length or a time, however the system (\ref{syst3}) has
soliton solutions and the width of these solitons can be considered as
a characteristic length, of order unity (in dimensioned units, it is
of order of the polarization healing length $\xi_p$). Nevertheless, if we
consider nonlinear structures at a much larger scale, as this is the
case for modulated waves whose envelopes change little over a
wavelength, then we can neglect such a `microscopic' length scale and
look for smooth solutions of the Whitham equations. In short: at the
`macroscopic' scale there is no characteristic length in the initial
conditions (\ref{init-cond}) and the solutions of the Whitham
equations can be sought as functions of the self-similar variable
$z=x/t$ only.

There exist also smooth solutions of the original system (\ref{syst3})
in which not the envelopes, but the functions $w(x,t)$ and $\,v(x,t)$
themselves depend slowly on the space coordinate. This corresponds to
a hydrodynamic regime where one can neglect the higher derivatives in
the second equation of the system (\ref{syst3}).  Again, such
hydrodynamic approximate solutions can only depend on the self-similar
variable $z=x/t$. These smooth non-dispersive waves can contribute
-- as DSWs do --- to the whole wave structures arising from the
space-time evolution of the initial profiles \eqref{init-cond}.

It is convenient, in a first stage, to select particular initial
conditions for which the time evolved wave structure reduces to a
single type of wave (hydrodynamic, modulated cnoidal, modulated
trigonometric, etc). In a second step (Sec. \ref{sec5}) we will
proceed to the full classification of the structures evolving from
arbitrary initial conditions, but in the present section we shall
first identify what we denote as ``key elements'', solutions of the
Riemann problem for specific values of the boundary value
constants. These are the building blocks of which are composed the
general self-similar solution of the Riemann problem. We shall start
with the hydrodynamic key elements which are solutions for which
dispersive effects can be neglected.

\subsection{Plateau and rarefaction waves}\label{sec4a}

As stated above, nonlinear polarization waves with typical length
scale much larger than unity can be described in the framework of a
dispersionless treatment in which the dispersive term in \eqref{syst3}
is omitted:
\begin{equation}
\begin{split}\label{nd1}
 w_t-[(1-w^2)v]_x=0\; ,\quad
 v_t-[(1-v^2)w]_x=0 \; .
\end{split}
\end{equation}
We shall denote these equations as {\it Bellevaux-Ovsyannikov
  equations} since they were first obtained independently by these
authors in the theory of two-layer shallow water dynamics
\cite{Lacombe65,Ovs79} (see also \cite{cavanie,sq-93,Lia00}).

First of all, we note that these equations have a simple solution
\begin{equation}\label{plat1}
  w(x,t)=\overline{w}=\mathrm{const}, \quad
v(x,t)=\overline{v}=\mathrm{const}.
\end{equation}
In spite of its `triviality', such a solution can play an important role
as an element of a self-similar structure, provided that its edge
points $x_{\pm}$ move with constant velocities $s_{\pm}=x_{\pm}/t$. We
shall call such an expanding region of constant flow a ``plateau
region''.

Now, since both variables $w$ and $v$ depend on the single variable
$z$, they can be considered as functions of each other and, hence,
such self-similar solutions are denoted as ``simple wave'' solutions
of the hydrodynamic equations (\ref{nd1}), see, e.g.,
Ref. \cite{Cou56}. For their study it is convenient to transform the
Bellevaux-Ovsyannikov system to a diagonal Riemann form by defining
the Riemann invariants \cite{cavanie}
\begin{equation}\label{nd2}
r_{\pm}=v w \pm \sqrt{(1-v^2)(1-w^2)}.
\end{equation}
As a result we obtain the system
\begin{equation}\label{nd3}
\partial_t r_\pm + v_\pm(r_-,r_+) \, \partial_x r_\pm =0\;,
\end{equation}
where $v_{\pm}$ are the ``Riemann velocities''
\begin{equation}\label{nd4}
\begin{split}
v_-=& \tfrac32 r_- + \tfrac12 r_+= 2 v w - \sqrt{(1-v^2)(1-w^2)}\; ,\\
v_+=&\tfrac12 r_- + \tfrac32 r_+=  2 v w + \sqrt{(1-v^2)(1-w^2)}\; .
\end{split}
\end{equation}
Eqs. \eqref{nd3} are reminiscent of the equations of compressible gas
dynamics , see, e.g., Refs. \cite{LL-6,Kam00}. We note here that,
although the relative density $w(x,t)=\cos\theta(x,t)$ is constrained
to vary between $-1$ and 1, the relative velocity $v(x,t)$ can assume,
generally speaking, any values (as will be exemplified in sections
\ref{sec5a} and \ref{sec5b}). However, in the regime we consider here,
because of the assumption of slow variation of the field $v(x,t)$, a
value larger than $1$ (or lower than $-1$) suffers from a dynamical
instability because it induces perturbations which grow exponentially
(as in the uniform case discussed in Sec.~\ref{sec2}), resulting in
oscillations which cannot be treated within the dispersionless
approximation. The Riemann variables \eqref{nd2} are thus always
properly defined only in the ``hyperbolicity region''
\begin{equation}\label{nd5}
  -1\leq v,w\leq 1,
\end{equation}
where the velocities (\ref{nd4}) are real.

One can also remark that the Riemann velocities
\eqref{nd4} expressed in terms of $v$ and $w$ correspond to the sound
velocity \eqref{vsound} for a uniform background characterized by $w$
and $v$, in agreement with the long wavelength approximation which is
at the heart of the dispersionless approximation.

For a simple wave solution, one of the Riemann invariants is constant,
and this condition (namely: either $r_-(v,w)=\mathrm{const}$ or
$r_+(v,w)=\mathrm{const}$) gives, when applied to Eq. \eqref{nd2}, the
above mentioned relationship between the variables $v$ and $w$.
Consequently, on the $(v,w)$-plane these simple wave solutions are
depicted as arcs of the ellipse $ (r_{\pm}-vw)^2=(1-v^2)(1-w^2) $ or
\begin{equation}\label{227.6}
  \frac{(v+w)^2}{2(1+r)}+\frac{(v-w)^2}{2(1-r)}=1,
\end{equation}
where $r$ denotes the constant value of $r_+$ or $r_-$.
This ellipse is inscribed into a square $-1\leq w,v\leq 1$ (domain of
hyperbolicity: cf.~(\ref{nd5})) and touches its sides at 4 points
with coordinates
\begin{equation}\label{endpoint}
(1,r), \; (-1,-r),\; (-r,-1)\;\;\mbox{and}\;\; (r,1)\; .
\end{equation}
If $r_-+r_+\geq 0$ (i.e. when $w$ and $v$ have the same sign), the
physical variables are expressed in terms of the Riemann invariants by the
formulas
\begin{equation}\label{227.8a}
\begin{split}
  &w=\pm\sqrt{\frac12\left[1+r_-r_+\pm\sqrt{(1-r_-^2)(1-r_+^2)}\right]},\\
  &v=\pm\sqrt{\frac12\left[1+r_-r_+\mp\sqrt{(1-r_-^2)(1-r_+^2)}\right]}.
  \end{split}
\end{equation}
Otherwise, for $r_-+r_+\leq 0$ (i.e. when $w$ and
$v$ have opposite signs)
\begin{equation}\label{227.8b}
\begin{split}
  &w=\pm\sqrt{\frac12\left[1+r_-r_+\pm\sqrt{(1-r_-^2)(1-r_+^2)}\right]},\\
  &v=\mp\sqrt{\frac12\left[1+r_-r_+\mp\sqrt{(1-r_-^2)(1-r_+^2)}\right]}.
  \end{split}
\end{equation}
It is clear that in the expressions \eqref{227.8a} and \eqref{227.8b}
one should have $|r_{\pm}|\leq 1$.

\begin{figure}
  \centering
  \includegraphics[width=\linewidth]{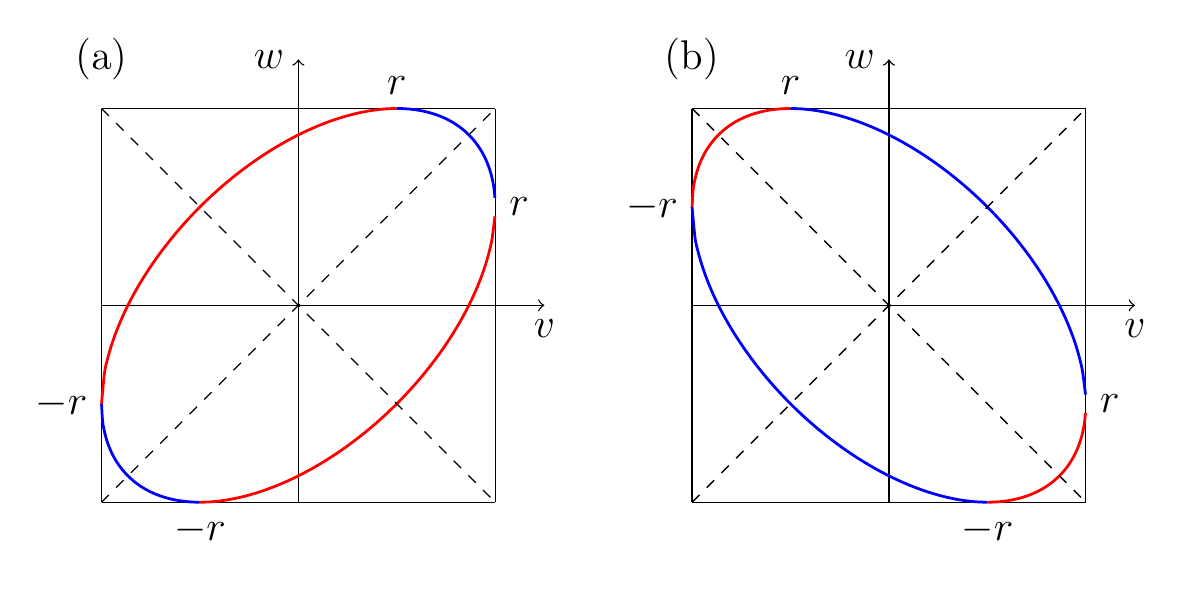}
  \caption{Simple-wave solutions in the $(v,w)$-plane. Along these
    branches of ellipse only one Riemann invariant varies, the value
    of the constant other one is denoted by $r$. In (a) we have $r>0$
    and in (b) $r<0$. Red arcs correspond to $r_+=\mathrm{const}$ and
    blue ones to $r_-=\mathrm{const}$.}
    \label{fig2}
\end{figure}

On the ellipses \eqref{227.6} we can express $w$ as a function of $v$
in an explicit form. To do so, it is convenient to distinguish four
arcs and to write
\begin{equation}\label{227.9}
\begin{split}
&w(v)=\\
&\left\{
  \begin{array}{l}
  r_+v+\sqrt{(1-r_+^2)(1-v^2)}, \quad -1\leq v\leq r_+=\mathrm{C}^{\rm st},\\
  r_+v-\sqrt{(1-r_+^2)(1-v^2)},\quad  -r_+=\mathrm{C}^{\rm st}\leq v\leq 1,\\
  r_-v-\sqrt{(1-r_-^2)(1-v^2)},\quad  -1\leq v\leq -r_-=\mathrm{C}^{\rm st},\\
  r_-v+\sqrt{(1-r_-^2)(1-v^2)},\quad  r_-=\mathrm{C}^{\rm st}\leq v\leq 1.
  \end{array}\right.
  \end{split}
\end{equation}
In Fig.~\ref{fig2} the arcs for constant $r_-$ are shown in blue and
those for constant $r_+$ in red.  The value of the constant is denoted
as $r$. On these arcs, the Riemann invariant which varies reaches its
maximal value (equal to 1) on the diagonal $w=v$ (for $r_+$) or on the
anti-diagonal $w=-v$ (for $r_-$); at the end points --- whose
coordinates are listed in \eqref{endpoint} --- is it equal to $r$.

\begin{figure}
  \centering
  \includegraphics[width=\linewidth]{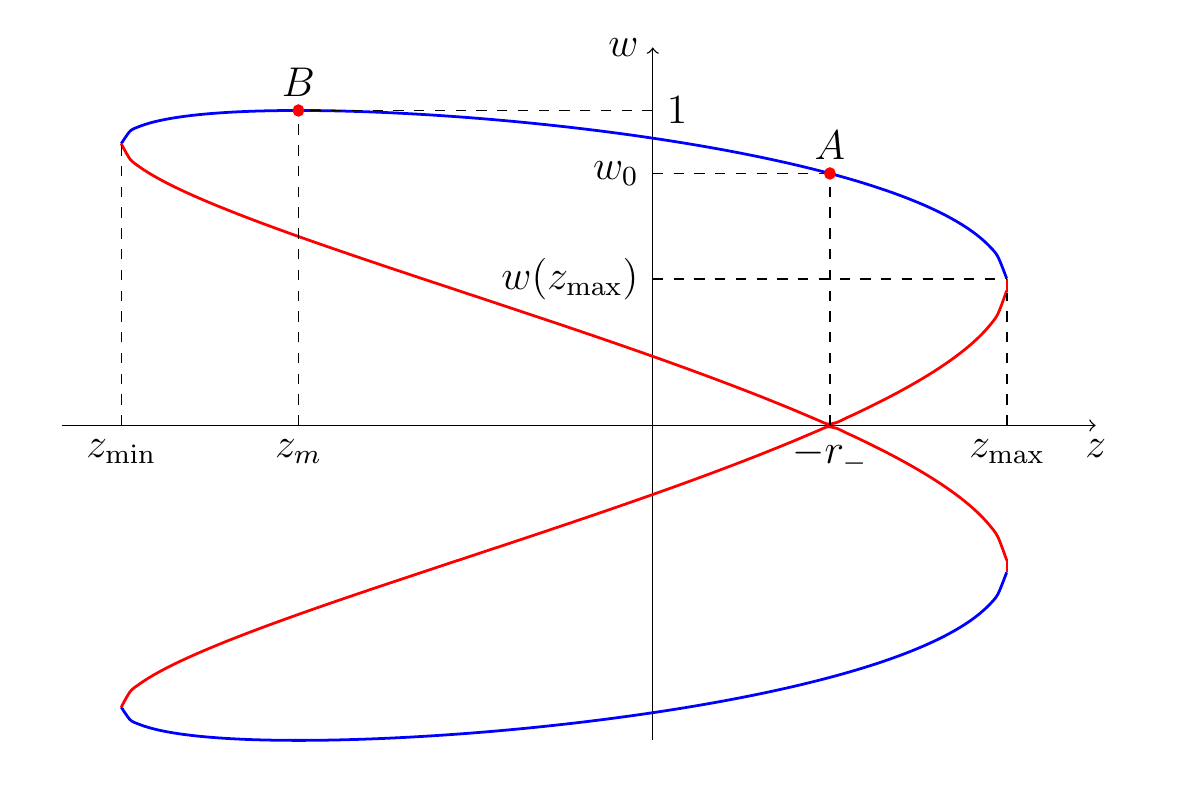}
  \caption{Dispersionless simple-wave solutions $w(z)$ plotted as functions of
    $z=x/t$. The figure is drawn for $r_-=\mathrm{const}$. This constant
    value is expressed by means of \eqref{nd2} in terms of two
    constants, $w_0>0$ and $v_0=0$, i.e., $r_-=-\sqrt{1-w_0^2}<0$ and
    $w_0=\sqrt{1-r_{-}^2}$. Here $z_{\mathrm{max}} =
    \tfrac12r_{-}+\tfrac32$, $w(z_{\mathrm{max}}) =
    \sqrt{(1+r_{-})/2}$, $z_{\mathrm{min}} = \tfrac12r_{-}-\tfrac32$
    and $w(z_{\mathrm{min}}) = \sqrt{(1-r_{-})/2}$. The value $w=1$ is
    reached for $z= z_m=2r_{-}= -2\sqrt{1-w_0^2}$. The arc between
    points $A$ and $B$ is used below for constructing an ``expansion
    into vacuum'' solution, see Fig. \ref{fig4}.}
\label{fig3}
\end{figure}

The dependence of $w$ and $v$ on the self-similar variable $z=x/t$ is
found at once by noticing that in this case the system (\ref{nd3})
reduces to
\begin{equation}\label{nd3b}
  (v_--z)\cdot\frac{dr_-}{dz}=0, \qquad (v_+-z)\cdot\frac{dr_+}{dz}=0.
\end{equation}
Hence, if one of Riemann invariants is constant, the Riemann velocity
of the other must be equal to $z$. Thus we arrive at two possible
solutions $r_-=\mathrm{const}$, $v_+=x/t\equiv z$ and
$r_+=\mathrm{const}$, $v_-=x/t\equiv z$. Let us consider the first one
in some detail. It is characterized by the relations
\begin{equation}\label{nd3c}
  v_+=\frac12r_-+\frac32r_+=z=\frac{x}t,\quad r_-=\mathrm{const}.
\end{equation}
 More explicitly, we have two equations
\begin{equation}\nonumber
\begin{split}
&v_+=2vw+\sqrt{(1-v^2)(1-w^2)}=z,\\
&r_-=vw-\sqrt{(1-v^2)(1-w^2)}=\mathrm{const},
\end{split}
\end{equation}
which yield
\begin{equation}\label{229.19}
\begin{split}
  &w(z)=\pm\Bigg\{\frac16\Bigg[3+2r_{-}z-r_{-}^2\\
  &\pm 2
\sqrt{(1-r_{-}^2)(z-\tfrac12r_{-}+\tfrac32)
(\tfrac12r_{-}+\tfrac32-z)}\Bigg]
\Bigg\}^{1/2},
  \end{split}
\end{equation}
and
\begin{equation}\label{229.19b}
  v(z)=\frac{r_{-}+z}{3\,w(z)}.
\end{equation}
The solution has thus four branches --- corresponding to four
possible choices of signs in (\ref{229.19}) --- that are located within
the interval
\begin{equation}\label{229.20}
  \frac12r_{-}-\frac32\leq z\leq \frac12r_{-}+\frac32 .
\end{equation}
Comparing \eqref{229.20} and \eqref{nd3c} one sees that at the end
points we have $r_+=\pm1$. The four branches of the solutions are
represented in Fig.~\ref{fig3}.  It is important to stress that the
solution expressed in terms of the Riemann invariants by
Eq.~(\ref{nd3c}) is mapped into four arcs in the $(v,w)$-plane and
four functions $w=w(z)$ given by Eq.~(\ref{229.19}).

Similar formulas and plots can be obtained for the solution
$r_+=\mathrm{const},v(r_-,r_+)=x/t\equiv z$.

\begin{figure}
  \centering
  \includegraphics[width=\linewidth]{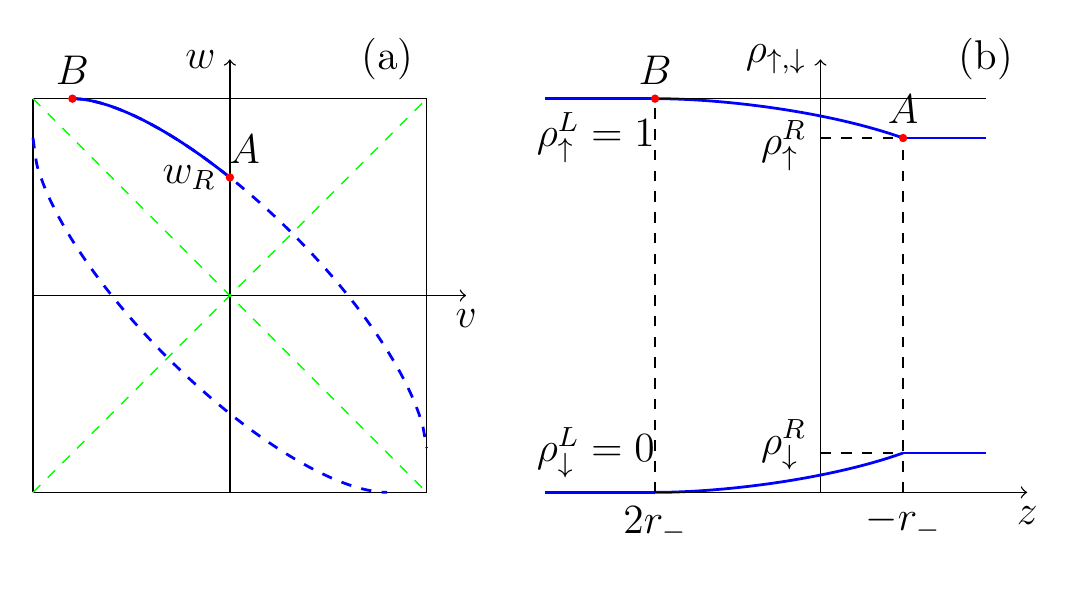}
  \caption{(a) The solid blue arc represents a simple-wave solution with
    $w_L=1$, $w_R>0$ and $v_R=v_L=0$ in the $(v,w)$-plane.  (b)
    Corresponding density profiles $\rho_{\uparrow,\downarrow}$
plotted as functions of $z=x/t$. }
\label{fig4}
\end{figure}

The simplest concrete situation of physical interest is represented in
Fig.~\ref{fig4}(a). It consists in
the path in the $(v,w)$ plane formed by a
single arc $AB$ which corresponds to a rarefaction wave shown in
Fig.~\ref{fig4}(b).  This arc is described by the last of formulae
~(\ref{227.9}).  Here the initial jump \eqref{init-cond} in the relative
density evolves
into a smooth rarefaction wave, similarly to what occurs in the `dam
break problem' in compressible fluid dynamics when a gas expands into vacuum
flowing along a tube after removal of a wall. At the initial state
both components are at rest, $v_L=v_R=0$, the total density is fixed,
i.e. $\rho_{\uparrow}+\rho_{\downarrow}=1$ everywhere and does not
change with time, and initially we have $\rho=\rho_{\uparrow}^L=1$ for
$x<0$, $\rho=\rho_{\uparrow}^R<1$ for $x>0$. This means that we have a
`vacuum' of the component $\rho_{\downarrow}$ for $x<0$ at the initial
time. The value of the constant Riemann invariant is fixed by
the parameters of the flow at point $A$:
$r_-=-\sqrt{1-w_R^2}=-2\sqrt{\rho_{\uparrow}^R(1-\rho_{\uparrow}^R)}$.
The parameters of the flow at the matching point
$A$ preserve their values during the time evolution and therefore
$r_+(A)=\sqrt{1-w_R^2}=-r_-$. Consequently, this edge of the
rarefaction wave propagates to the right at a velocity
$s_+=(r_-+3r_+(A))/2=-r_-=
\sqrt{1-w_R^2}=2\sqrt{\rho_{\uparrow}^R(1-\rho_{\uparrow}^R)}$, which
coincides with the polarization sound velocity (\ref{vsound}) in this
case.  At point $B$ we have $w(B)=1$, hence
$r_+(B)=v(B)=r_-=-\sqrt{1-w_R^2}$, and this edge propagates to the
left with velocity $s_-=(r_-+3r_+(B))/2=2r_-=
\sqrt{1-w_R^2}=-4\sqrt{\rho_{\uparrow}^R(1-\rho_{\uparrow}^R)}$.  As
we see, this is not the sound velocity of waves in the
component $\rho_{\uparrow}$, but rather the maximal velocity of
expansion of the component $\rho_{\downarrow}$ into its vacuum.  This
quasi-one-dimensional flow of two-component BEC was studied
numerically in Ref. \cite{Ham11} and analytically in Ref. \cite{Con16}.

\begin{figure}
  \centering
  \includegraphics[width=\linewidth]{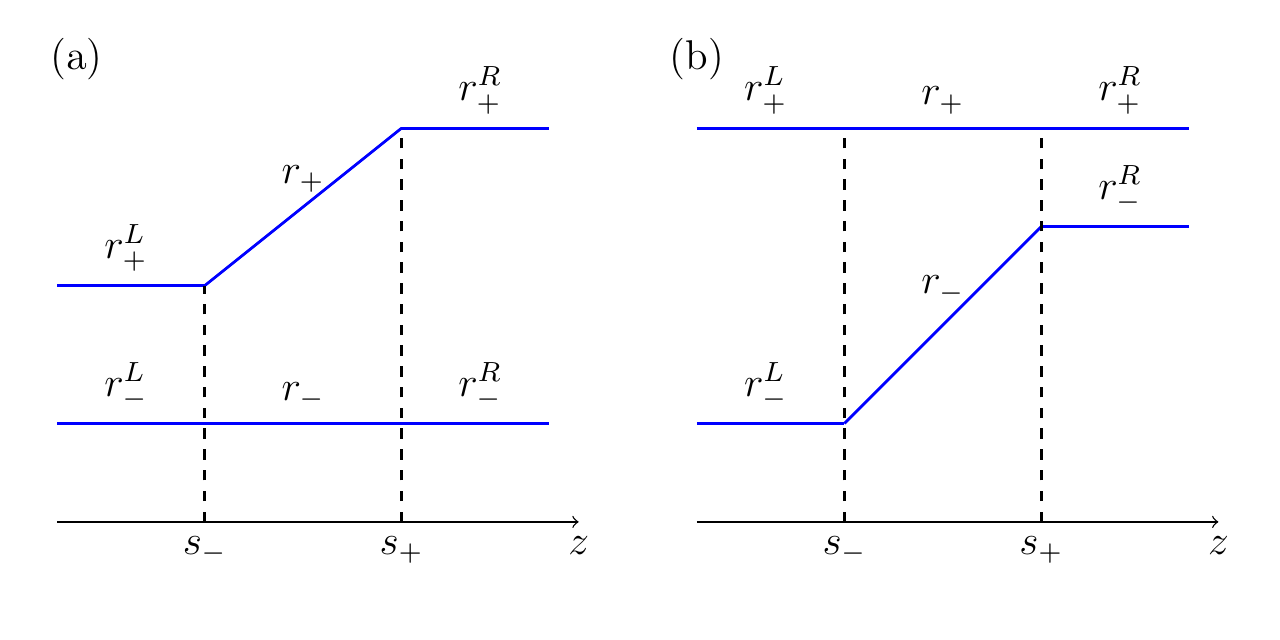}
  \caption{Sketches of the behavior of the Riemann invariants in
    rarefaction wave solutions of the hydrodynamic equations with
    boundary conditions (\ref{281.6}): (a) $r_-=\mathrm{const}$,
    $r_+^L<r_+^R$; (b) $r_+=\mathrm{const}$, $r_-^L<r_-^R$.}
\label{fig5}
\end{figure}

In the above solution, the rarefaction wave connects two plateaus with
parameters $v_L=r_-$, $w_L=1$ and $v_R=0$, $w_R=\sqrt{1-r_-^2}$, in such a
way that the Riemann invariant $r_-$ is constant along the wave and
the plateaus. It is clear that this solution can be generalized to any
rarefaction waves connecting two plateaus provided the following two
conditions are fulfilled. First, one of the Riemann invariants must
have the same value on both plateaus,
\begin{equation}\label{281.6}
  \text{(a)}\quad r_-^L=r_-^R \quad\text{or}\quad
\text{(b)}\quad r_+^L=r_+^R.
\end{equation}
Second, since the one of the Riemann invariants which varies is a
solution of type (\ref{nd3c}) which depends monotonously on $z$, the
dependence of the Riemann invariants in terms on the physical
parameters must also be monotonous in order to keep the solution
single-valued. This means that the two edge points of the rarefaction
wave must lie within one of the four triangles which are obtained by
cutting the hyperbolicity square by its diagonals along which the
Riemann invariants reach their extremal values. We shall denote these
triangles as {\it monotonicity triangles}. They play an important role
in the classification of the wave patterns because they define the
domains where the characteristic velocities (\ref{nd4}) satisfy the
conditions of genuine nonlinearity (see, e.g., Ref. \cite{lax}).
Besides that, both edge points must lie on the same branch of the
ellipse and should not be separated by a point where the ellipse
touches a side of the hyperbolicity square (the four sides of this
square correspond to $v=\pm 1$ and $w\in [-1,1]$ or $w=\pm 1$ and
$v\in [-1,1]$). For these rarefaction wave solutions, the behavior of
the Riemann invariants considered as functions of $z$ is displayed in
Fig.~\ref{fig5}.
The edge velocities of the rarefaction waves are equal to
\begin{equation}\label{sk1}
  \begin{split}
  & (a)\quad
s_-=\frac{1}{2}r_-^L+\frac{3}{2}r_+^L,
\quad s_+=\frac{1}{2}r_-^R+\frac{3}{2}r_+^R;\\
  & (b)\quad s_-=\frac{3}{2}r_-^L+\frac{1}{2}r_+^L,
\quad s_+=\frac{3}{2}r_-^R+\frac{1}{2}r_+^R.
  \end{split}
\end{equation}
As we shall see in the next sections, such rarefaction waves are
among the key elements from which a generic wave pattern may be
composed.  Obviously, their observation implies that $r_+^L<r_+^R$ in
case (a) or $r_-^L<r_-^R$ in case (b), which imposes conditions on the
parameters of the initial discontinuity. It is natural to ask what
happens if the boundary conditions correspond to opposite inequalities
(namely $r_+^L>r_+^R$ or $r_-^L>r_-^R$); this question leads us to the
study of another type of key elements --- dispersive shock waves.

\subsection{Cnoidal dispersive shock waves}\label{sec4b}

If we try naively to use a formal self-similar solution of the type
(\ref{nd3c}) for describing a wave satisfying boundary conditions
such that $r_+^L>r_+^R$ or $r_-^L>r_-^R$, then we arrive at once to
physically meaningless multi-valued solutions (see, e.g.,
\cite{Kam00}) which represent the simplest wave breaking situation. In
this case, the major insight of Gurevich and Pitaevskii
\cite{gp-73} has been to take into account the dispersive effects
which lead to the generation of oscillations in regions where the
physical variable have large spatial derivatives: the multi-valued
solution must be replaced by a modulated nonlinear periodic solution whose
parameters satisfy the Whitham equations (at least for large enough evolution
time). As a matter of fact,
this oscillating wave structure replaces the well-known
shock waves occurring in viscous compressible fluid dynamics and hence
it is called a {\it dispersive shock wave} (DSW).

\begin{figure}
  \centering
  \includegraphics[width=\linewidth]{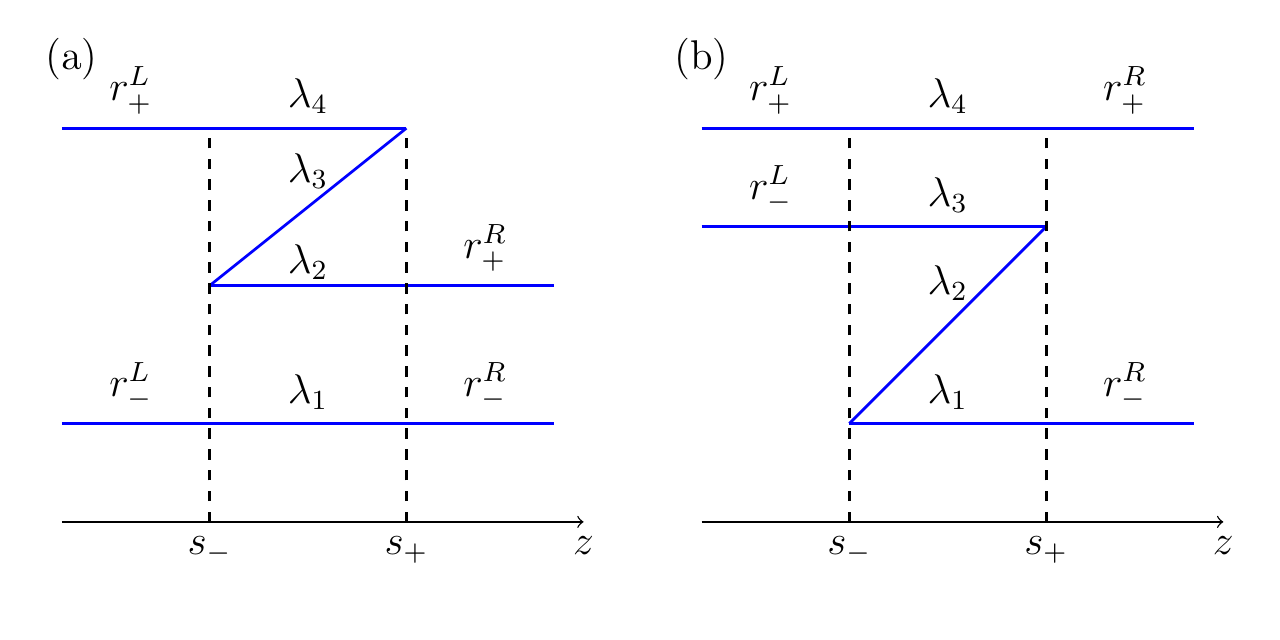}
  \caption{Sketches of the behavior of the Riemann invariants in DSW
    solutions of the Whitham equations with the boundary conditions
    (a) $r_-^L=r_-^R$, $r_+^L>r_+^R$ or (b) $r_+^L=r_+^R$,
    $r_-^L>r_-^R$.}
  \label{fig6}
\end{figure}

From a formal point of view, we look again for self-similar solutions,
here not for the equations \eqref{nd3}, but instead for
the Whitham equations (\ref{276.56}). Assuming in these equations that
the $\la$'s depend only on the
variable $z=x/t$ we obtain at once
\begin{equation}\label{282.1}
  \left\{v_i(\la)-z\right\}\cdot\frac{d\la_i}{dz}=0, \quad i=1,2,3,4.
\end{equation}
In the case where the Whitham velocities are given by
Eq.~(\ref{276.55}), one can satisfy this system if three Riemann
invariants remain constant while the fourth one varies in such a way
that the expression in the curly bracket vanishes. We assume that at
its both edges the DSW matches with a smooth solution of the
hydrodynamic equations and therefore at these edges the averaged
equations should reproduce the same dynamics as the dispersionless
hydrodynamic equations do. The comparison of the limiting expressions
(\ref{277.56})--(\ref{277.58}) with the diagonal form of the
hydrodynamic equations (\ref{nd3}), (\ref{nd4}) shows that the
matching conditions can be satisfied if the Riemann invariants behave
as represented in Fig.~\ref{fig6}. It is clear that wave structures of
this type appear only if the flows at the edges of the DSW satisfy
either the condition (a) $r_-^L=r_-^R$ or (b) $r_+^L=r_+^R$ which
coincide with (\ref{281.6}).  According to Fig.~\ref{fig6}, the three
constant Riemann invariants in the solutions of the Whitham equations
are determined by the boundary conditions and the $z$-dependence of the
remaining one is determined by the vanishing of the expression
in curly brackets in Eqs.~(\ref{282.1}):
\begin{equation}\label{282.4}
  \begin{split}
 (a)& \quad \la_1=r_-^L\; , \quad \la_2=r_+^R\; ,\quad \la_4=r_+^L\; ,\\
  & \quad v_3(r_-^L,r_+^R\; ,\la_3(z),r_+^L)=z\; ;\\
  (b)& \quad \la_1=r_-^R\; , \quad \la_3=r_-^L\; ,\quad \la_4=r_+^L\; ,\\
 & \quad v_2(r_-^R,\la_2(z),r_-^L,r_+^L)=z\; .
  \end{split}
\end{equation}
These formulas show that the edge velocities of the DSW are
unambiguously determined by the values of the Riemann invariants at
its boundaries. They are equal to
\begin{equation}\label{sk2}
  \begin{split}
  (a) & \quad s_-=\frac{1}{2}(r_-^L+2r_+^R+r_+^L)\; , \\
& \quad s_+=2r_+^L+\frac{(r_+^R-r_-^R)^2}{2(r_+^R+r_-^R-2r_+^L)}\; ;\\
  (b) & \quad s_-=2r_+^R+\frac{(r_+^L-r_-^L)^2}{2(r_+^L+r_-^L-2r_+^R)}\; ,\\
& \quad s_+=\frac{1}{2}(r_-^R+2r_+^L+r_+^R)\; .
  \end{split}
\end{equation}
However, the situation changes in what concerns the envelopes of the DSWs,
because the mapping (\ref{272.30}) of the $\la$'s to the
physical parameters $(v,w)$ is multi-valued. As a result,
each of the $\la$-diagrams in Fig.~\ref{fig6} (a) or (b) corresponds
to four different DSWs. To clearly see this,
let us consider the limiting expressions of
Eqs.~(\ref{272.30}) at the edges of the DSW.

We first assume that $f_1>0$ is given by Eq.~(\ref{270.16a}). Then,
after some calculations, we obtain for the case of Fig.~\ref{fig6}(a),
at the soliton edge with $\la_3=\la_2$, the expressions
\begin{subequations}\label{273.31}
\begin{equation}\label{273.31a}
  w_1=-\sqrt{\tfrac12\left[1+(2\la_2^2-1)S^{(+)}_{1,4}-2\la_2\la_2'
E^{(-)}_{1,4}\right]},
\end{equation}
\begin{equation}\label{273.31b}
  w_2=w_3=-\sqrt{\tfrac12(1+\la_1\la_4-\la_1\la_4')},
\end{equation}
\begin{equation}\label{273.31c}
  w_4=\sqrt{\frac12\left[1+(2\la_2^2-1)S^{(+)}_{1,4}
+2\la_2\la_2'E^{(-)}_{1,4}\right]},
\end{equation}
\end{subequations}
where, for shortening the formulas, we have introduced the notations
\begin{equation}\label{SandE}
S^{(\pm)}_{i,j}=\lambda_i\lambda_j\pm\lambda'_i\lambda'_j,
\quad\mbox{and}\quad
E^{(\pm)}_{i,j}=\lambda_i\lambda_j'\pm\lambda'_i\lambda_j.
\end{equation}
At the small amplitude edge with $\la_3=\la_4$ one obtains
\begin{subequations}\label{273.32}
\begin{equation}\label{273.32a}
  w_1=-\sqrt{\tfrac12\left[1+(2\la_4^2-1)S^{(+)}_{1,2}-
2\la_4\la_4'E^{(-)}_{1,2}\right]},
\end{equation}
\begin{equation}\label{273.32b}
  w_2=-\sqrt{\tfrac12\left[1+(2\la_4^2-1)S^{(+)}_{1,2}
+2\la_4\la_4'E^{(-)}_{1,2}\right]},
\end{equation}
\begin{equation}\label{273.32c}
  w_3=w_4=\sqrt{\tfrac12(1+\la_1\la_2-\la_1\la_2')}.
\end{equation}
\end{subequations}
If we change the sign of $f_1$, then, for $f_1<0$, these expressions will
also change sign with appropriate reordering.

In a similar way, for the case of Fig.~\ref{fig6}(a) and when $f_1>0$ is
given by Eq.~(\ref{270.16b}), we obtain at the soliton edge with
$\la_3=\la_2$ the expressions
\begin{subequations}\label{273.35}
\begin{equation}\label{273.35a}
  w_1=-\sqrt{\tfrac12\left[1+(2\la_2^2-1)S_{1,4}^{(-)}+2\la_2\la_2'
E_{1,4}^{(+)}\right]},
\end{equation}
\begin{equation}\label{273.35b}
  w_2=w_3=-\sqrt{\tfrac12(1+\la_1\la_4+\la_1\la_4')},
\end{equation}
\begin{equation}\label{273.35c}
  w_4=\sqrt{\tfrac12\left[1+(2\la_2^2-1)S_{1,4}^{(-)}
-2\la_2\la_2'E_{1,4}^{(+)}\right]},
\end{equation}
\end{subequations}
and at the small amplitude edge with $\la_3=\la_4$ the expressions
\begin{subequations}\label{273.36}
\begin{equation}\label{273.36a}
  w_1=w_2=-\sqrt{\tfrac12(1+\la_1\la_2+\la_1\la_2'))}.
\end{equation}
\begin{equation}\label{273.36b}
  w_3=-\sqrt{\tfrac12\left[1+(2\la_4^2-1)S_{1,2}^{(-)}+
2\la_4\la_4'E_{1,2}^{(+)}\right]},
\end{equation}
\begin{equation}\label{273.36c}
  w_4=\sqrt{\tfrac12\left[1+(2\la_4^2-1)S_{1,2}^{(-)}
-2\la_4\la_4'E_{1,2}^{(+)}\right]}.
\end{equation}
\end{subequations}
Again, if we change the sign of $f_1$ then, for $f_1<0$, these
expressions also change signs with appropriate reordering.

We now consider the diagram of Fig.~\ref{fig6}(b) and assume that $f_1>0$
is given by Eq.~(\ref{270.16a}). Then we obtain at the soliton edge
with $\la_3=\la_2$ the expressions
\begin{subequations}\label{273.34}
\begin{equation}\label{273.34a}
  w_1=\sqrt{\tfrac12\left[1+(2\la_3^2-1)S_{1,4}^{(+)}
-2\la_3\la_3'E_{1,4}^{(-)}\right]},
\end{equation}
\begin{equation}\label{273.34b}
  w_2=w_3=\sqrt{\tfrac12(1+\la_1\la_4-\la_1\la_4')},
\end{equation}
\begin{equation}\label{273.34c}
  w_4=\sqrt{\tfrac12\left[1+(2\la_3^2-1)S_{1,4}^{(+)}
+2\la_3\la_3'E_{1,4}^{(-)}\right]},
\end{equation}
\end{subequations}
and at the small amplitude edge with $\la_2=\la_1$ the expressions
\begin{subequations}\label{273.33}
\begin{equation}\label{273.33a}
  w_1=w_2=-\sqrt{\tfrac12(1+\la_3\la_4-\la_3\la_4')},
\end{equation}
\begin{equation}\label{273.33b}
  w_3=\sqrt{\tfrac12\left[1+(2\la_1^2-1)S_{3,4}^{(+)}
-2\la_1\la_1'E_{3,4}^{(-)}\right]},
\end{equation}
\begin{equation}\label{273.33c}
  w_4=\sqrt{\tfrac12\left[1+(2\la_1^2-1)S_{3,4}^{(+)}
+2\la_1\la_1'E_{3,4}^{(-)}\right]}.
\end{equation}
\end{subequations}
If we take $f_1<0$ then these expressions will also change their signs with
appropriate reordering.

At last, for the case of Fig.~\ref{fig6}(b), when $f_1>0$ is given by
Eq.~(\ref{270.16b}), we obtain at the soliton edge with $\la_3=\la_2$
the expressions
\begin{subequations}\label{274.38}
\begin{equation}\label{274.38a}
  w_1=-\sqrt{\tfrac12\left[1+(2\la_3^2-1)S_{1,4}^{(-)}
+2\la_3\la_3'E_{1,4}^{(+)}\right]},
\end{equation}
\begin{equation}\label{274.38b}
  w_2=w_3=-\sqrt{\tfrac12(1+\la_1\la_4+\la_1\la_4')},
\end{equation}
\begin{equation}\label{274.38c}
  w_4=\sqrt{\tfrac12\left[1+(2\la_3^2-1)S_{1,4}^{(-)}
-2\la_3\la_3'E_{1,4}^{(+)}\right]},
\end{equation}
\end{subequations}
and at the small amplitude edge with $\la_2=\la_1$ the expressions
\begin{subequations}\label{274.37}
\begin{equation}\label{274.37a}
  w_1=w_2=-\sqrt{\tfrac12(1+\la_3\la_4+\la_3\la_4')}.
\end{equation}
\begin{equation}\label{274.37b}
  w_3=-\sqrt{\tfrac12\left[1+(2\la_1^2-1)S_{3,4}^{(-)}+
2\la_1\la_1'E_{3,4}^{(+)}\right]},
\end{equation}
\begin{equation}\label{274.37c}
  w_4=\sqrt{\tfrac12\left[1+(2\la_1^2-1)S_{3,4}^{(-)}
-2\la_1\la_1'E_{3,4}^{(+)}\right]}.
\end{equation}
\end{subequations}
If $f_1<0$ then these expressions will also change their signs with
appropriate reordering.

Thus we see indeed that each diagram in Fig.~\ref{fig6} corresponds to
four different sets of values for the $w_i$'s. It is important to
notice that for each set, the edges of these DSWs match with plateaus
and assume limiting values coinciding with the dispersionless
expressions (\ref{227.8a}) or (\ref{227.8b}). To avoid possible
confusion, it is worth noticing that the above limiting expressions
are correct not only for the self-similar situation but also for the
general case schematically represented in Fig.~\ref{fig1}.

\begin{figure}
  \centering
  \includegraphics[width=0.8\linewidth]{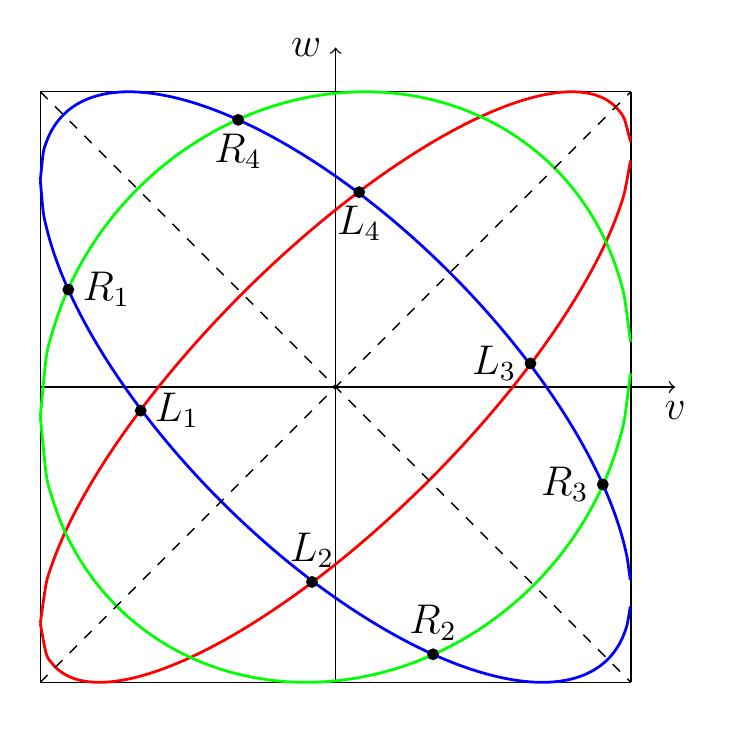}
  \caption{Plots of the ellipses of the $(v,w)$-plane along which the
    Riemann invariants $r_-^L=r_-^R=-0.7$ (blue), $r_+^L=0.8$ (red)
    and $r_+^R=0.1$ (green) are constant. Their crossing points define
    possible values of $(v,w)$ at the edges of the DSWs.}
  \label{fig7}
\end{figure}
It is convenient to symbolize the occurrence of a DSW by a diagram in
the $(v,w)$-plane. Let us consider for instance a possible DSW
corresponding to Fig.\ref{fig6}(a). The equal Riemann invariants
$r_-^L=r_-^R$ both correspond to the ellipse (\ref{227.6}) represented
in Fig.~\ref{fig7} by a blue line. Its intercepts with the ellipse
corresponding to the constant $r_+^L$ --- shown in red --- represent
possible values of $v_L$ and $w_L$ at the left (soliton) edge; its
intercepts with the ellipse corresponding to the constant $r_+^R$ --
shown in green --- represent possible values of $v_R$ and $w_R$ at the
right (small amplitude) edge. As we see in the figure, we get four
possible pairs of boundary conditions leading to cnoidal dispersive
shocks having all the same edge velocities but describing different
physical situations. In particular, $w_L$ in $L_1$ is given by
Eq.~(\ref{273.31b}) with $\la_1=r_-^L=r_-^R$, $\la_4=r_+^L$, and $w_R$
in $R_1$ by Eq.~(\ref{273.32c}) with $\la_1=r_-^L=r_-^R$,
$\la_2=r_+^L$. It is important to notice that each pair of boundary
points ($L_3$ and $R_3$, say) is located within a triangle obtained by
cutting the hyperbolicity square by its diagonals. It means that a
cnoidal DSW is possible only if both its edge points belong to the
same monotonicity triangle (region of genuine nonlinearity, earlier
defined in Sec. \ref{sec4a}). Since the edge points of the DSW belong
to an ellipse of constant Riemann invariant, we can {\it
  schematically} represent each DSW by an arc of this ellipse in the
$(v,w)$ plane. But we should keep in mind that --- at variance with the
dispersionless situation --- the actual plot representing how $v$ and
$w$ evolve within a DSW displays large oscillations and noticeably
departs from this ellipse, with which it has only the edge points in
common.

\begin{figure}
  \centering
  \includegraphics[width=\linewidth]{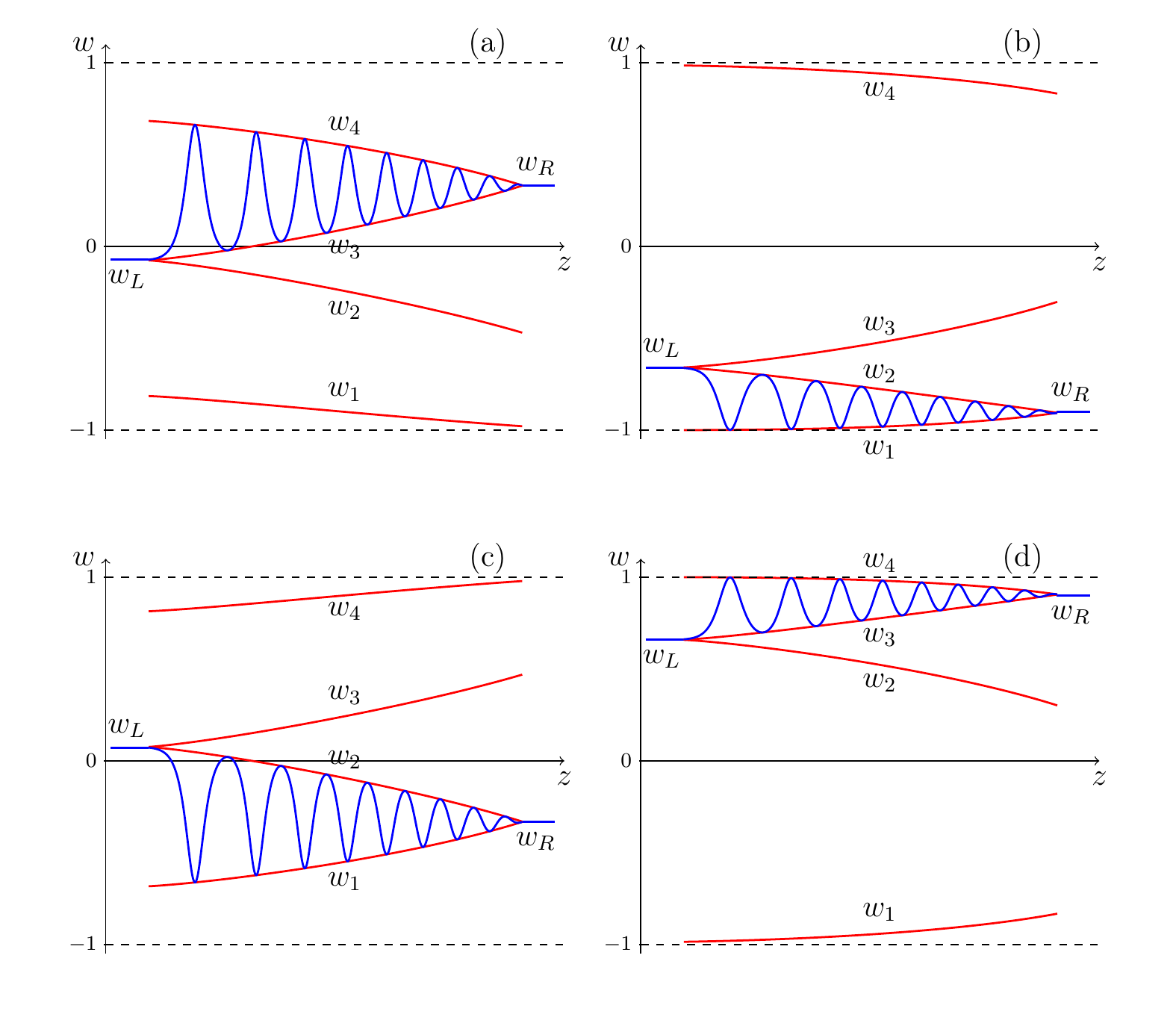}
  \caption{Plots of the functions $w_i(z)$ (red) and of the associated
    dispersive shock waves (blue) corresponding to the diagram
    Fig.~\ref{fig6}(a) and to the four possible choices of $f_1$ in
    Eqs.~(\ref{270.16}). In each case, two of the $w_i(z)$'s are the
    envelopes of the oscillatory structure, either $w_1$ and $w_2$ or
    $w_3$ and $w_4$, as clear from Eqs. \eqref{eq20} and
    \eqref{eq30}.}
  \label{fig8}
\end{figure}

\begin{figure}
  \centering
  \includegraphics[width=\linewidth]{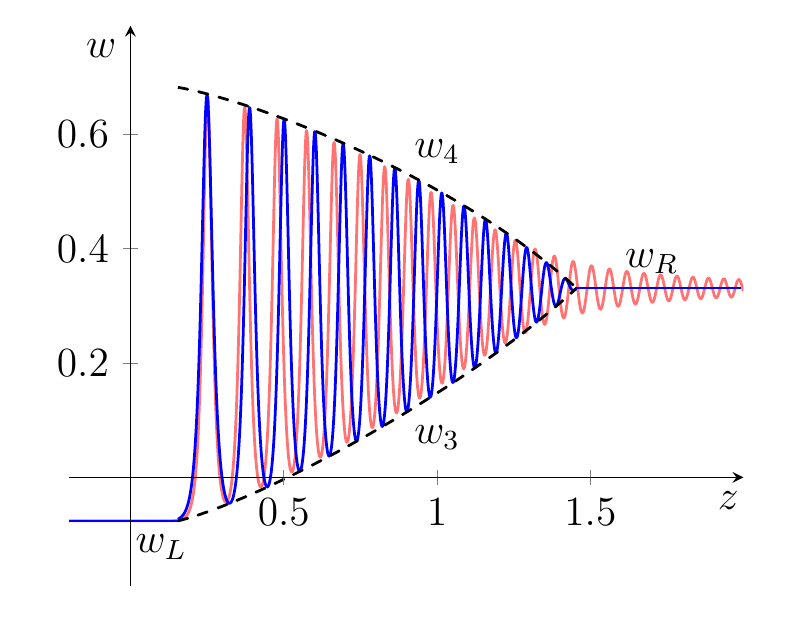}
  \caption{Dispersive shock wave evolving from an initial
    discontinuity with parameters corresponding to the points
    $L_1,R_1$ in Fig.~\ref{fig7}, with $r_-^L=r_-^R=-0.7$,
    $r_+^L=0.8$, $r_+^R=0.1$, which corresponds to $v_L=-0.659$,
    $w_L=-0.076$, $v_R=-0.906$ and $w_R=0.331$. The edge velocities
    are equal to $s_-=0.15$ and $s_+=1.45$. The analytic solution
    determined within the Whitham approximation scheme is shown by a
    blue line and the corresponding envelope functions by dashed black
    lines. The numerical solution computed for an evolution time
    $t=100$ is shown by a red line. According to our definition
    $v=(v_\downarrow - v_\uparrow)/(2c_p)$, and the wave structure with
    these parameters propagates to the right.}
  \label{fig9}
\end{figure}

The substitution of the solutions (\ref{282.4}) into (\ref{272.30})
gives the dependence of the $w_i$'s in term of $z$. Since we have four
sets of formulas corresponding to the four different choices of $f_1$
[Eqs. \eqref{270.16a} and \eqref{270.16b}], each of the two solutions
(a) and (b) in (\ref{282.4}) corresponds to four possible oscillatory
behaviors for the DSW.  The plots of the functions $w_i(z)$
produced by the diagram Fig.~\ref{fig6}(a), are shown in
Fig.~\ref{fig8}: cases (a) and (b) correspond to the positive signs in
Eqs.~(\ref{270.16}) and to arcs $L_1R_1$ and $L_2R_2$ in
Fig.~\ref{fig7}; cases (c) and (d) correspond to the negative signs in
Eqs.~(\ref{270.16}) and to arcs $L_3R_3$ and $L_4R_4$ in
Fig.~\ref{fig7}. Obviously, the plots \ref{fig8}(c) and \ref{fig8}(d)
can be obtained from the plots \ref{fig8}(a) and \ref{fig8}(b) by the
transformation $w\to-w$. It is worth noticing that if we exchange the
left and right boundary conditions, then the time evolution of the
initial flow yields to the formation, not of a DSW, but
of a rarefaction wave, such as
considered in the previous subsection. In Fig.~\ref{fig9} we compare
the analytic solution in the Whitham approximation with the exact
numerical solution of the Landau-Lifshitz system for the case shown in
Fig.~\ref{fig8}(a), with $v_L=-0.659$, $w_L=-0.076$, $v_R=-0.906$ and
$w_R=0.331$, which corresponds to $r_-^L=r_-^R=-0.7$, $r_+^L=0.8$ and
$r_+^R=0.1$. One can see that the envelope functions resulting from
the Whitham approach (dashed lines) agree very well with the exact
numerical solution.

In a similar way, the diagram Fig.~\ref{fig6}(b) with $r_+^L=r_+^R$
produces four other wave structures which correspond to four arcs
connecting the crossing points of the red and green ellipses in
Fig.~\ref{fig7}. Since this case does not differ essentially from the
above presented one, we shall not discuss it further.

The DSWs studied in the present subsection, as the rarefaction waves
presented in Sec.~\ref{sec4a}, can serve as key elements involved in
the description of a general wave structure evolving from the initial
conditions \eqref{init-cond}. They can be observed alone, in their
genuine form, only if the points corresponding to the left and right
boundaries belong to the same triangle of monotonicity. The
transitions between two triangles imply one more element, {\it contact
dispersive shocks}, and related structures which we consider in the
next section.

\subsection{Contact dispersive shock waves}\label{sec4c}

\begin{figure}
  \centering
  \includegraphics[width=0.75\linewidth]{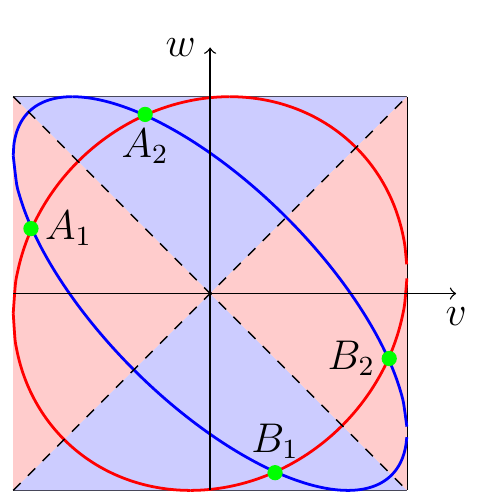}
  \caption{Plots of the ellipses of the $(v,w)$-plane along which the Riemann
    invariants $r_-^L=r_-^R=-0.7$ (red) and $r_+^L=r_+^R=0.1$ (blue)
    are constant. Their crossing points define possible values
    of $(v,w)$ at the edges of the contact dispersive shock.}
  \label{fig10}
\end{figure}
We now turn to the study of the situation where the left and right
boundaries points belong to different monotonicity triangles. In this
case the problem is no longer genuinely nonlinear. We shall start by
studying the simplest possible configuration in which the Riemann
invariants have equal values at both edges of the shock, i.e., when
$r_-^L=r_-^R$, and $r_+^L=r_+^R$. This situation resembles the one of
the so called `contact discontinuities' which play an important role
in the theory of viscous shocks (see, e.g., Ref. \cite{LL-6}); therefore we
shall denote the wave structures arising in this case as {\it contact
  dispersive shock waves} (CDSW).  (To avoid any confusion, we should
mention that in the dynamics of immiscible condensates, interfaces
between two components may appear which play the same role as the one
played by contact discontinuities in the theory of viscous shocks;
see, e.g., Ref. \cite{ik-2017}.)  In the case of a CDSW the ``left'' and
``right'' ellipses of constant Riemann invariants in the $(v,w)$-plane
coincide with each other, cf.~Fig.~\ref{fig10}. The intersections of
the ellipses $r_-^{(L,R)}= \mathrm{const}$ and $r_+^{(L,R)}=
\mathrm{const}$ define four points denoted as $A_1$, $A_2$, $B_1$ and
$B_2$ in Fig. \ref{fig10}.  These points can refer to either the left
or the right edge depending on the choice of $f_1$.

\begin{figure}
  \centering
  \includegraphics[width=9cm]{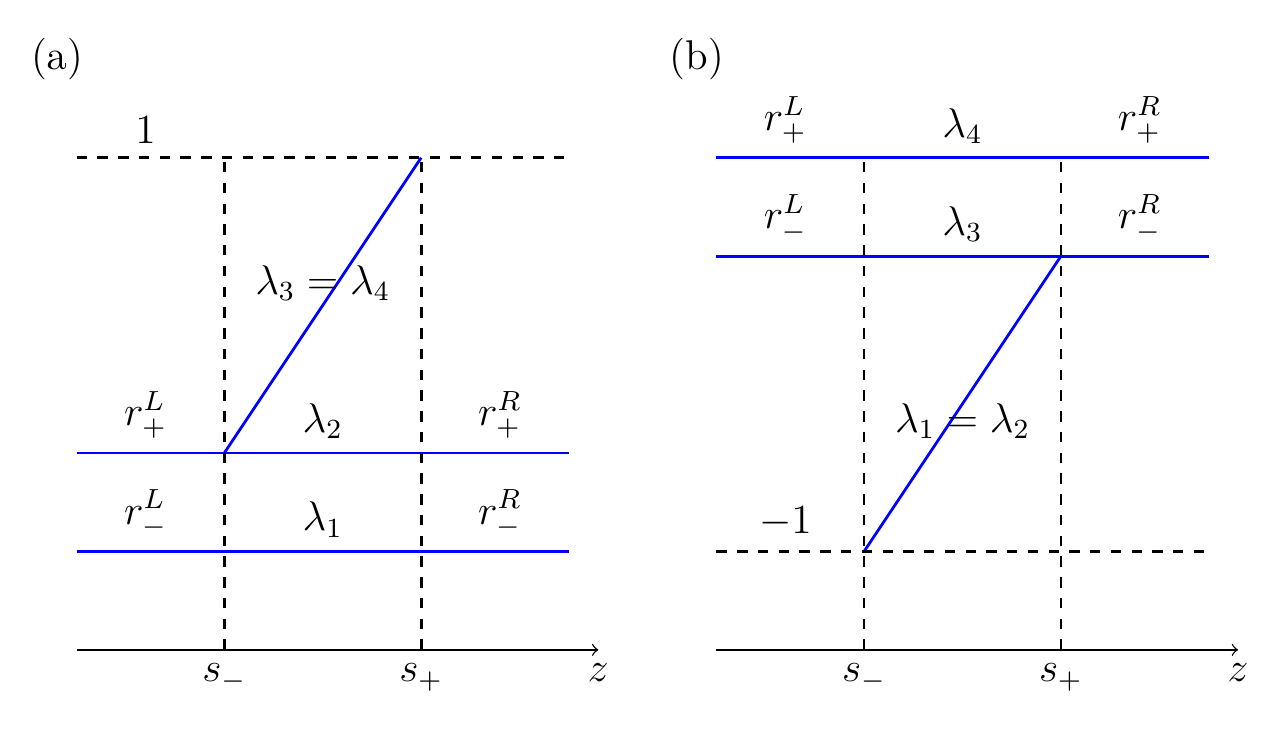}
  \caption{Sketches of the behavior of the Riemann invariants in
    contact dispersive shock wave solutions of the Whitham equations
    with the boundary conditions (a) $r_-^L=r_-^R$ or (b)
    $r_+^L=r_+^R$.}
  \label{fig11}
\end{figure}

First of all, we should determine the generic behavior of the Riemann
invariants in the case of interest here, and draw diagrams
representing the solutions of the Whitham equations
equivalent to the ones displayed in Figs.~\ref{fig5} and
\ref{fig6}. To be definite, let us consider the example represented in
Fig.~\ref{fig10}, with a left edge corresponding to point $A_1$, and a
right one to $B_1$.  In this case, the arc of ellipse connecting the
end points crosses the main diagonal $w=v$ of the hyperbolicity square
along which one of the dispersionless Riemann invariants takes its
maximal value, equal to unity: $r_+=1$. This means that in the formal
dispersionless solution, the invariant $r_+$ would first increase and
reach its maximal value $r_+=1$, then decrease down to the initial
value $r_+^R=r_+^L$ along the same `path' $r_+=\tfrac23(z-2r_-^L)$
[cf.~Eq.~\eqref{nd3c}]. By analogy with the case of a regular cnoidal
shock considered in the preceding subsection, it is natural to assume
that the actual behavior of the Riemann invariants $\la_i$ corresponding to
the Whitham equations reproduces here also the same qualitative
structure as the one expected on the basis of the dispersionless
analysis. This leads in the present case to the situation depicted in
Fig.~\ref{fig11}(a), where the invariants $\la_1$ and $\la_2$ remain
constant within the shock region (and match the boundary conditions:
$\la_1=r_-^L=r_-^R$, $\la_2=r_+^L=r_+^R$), whereas the two other
Riemann invariants are equal ($\la_3=\la_4$) and satisfy the same
Whitham equation $v_3(r_-^L,r_+^L, \la_4,\la_4) = v_4(r_-^L,r_+^L,
\la_4,\la_4)=z$ [with $v_3$ and $v_4$ given by the appropriate version
of Eq.~(\ref{277.57})]. We thus get
\begin{equation}\label{284.1}
\begin{split}
 &\la_1=r_-^L=r_-^R,\quad \la_2=r_+^L=r_+^R,\\
 & v_3=v_4=2\la_4+\frac{(r_+^L-r_-^L)^2}{2(r_+^L+r_-^L-2\la_4)}=z,
 \end{split}
\end{equation}
where the last formula determines the dependence of $\la_4$ on $z$,
which can be presented in an explicit form
\begin{equation}\label{285.1}
\begin{split}
  \la_4(z)=\frac14
& \bigg[z+r_+^L+r_-^L+\\
& \sqrt{(z-r_+^L-r_-^L)^2+2(r_+^L-r_-^L)^2}\bigg].
\end{split}
\end{equation}
Here $z$ varies within the interval $s_-\leq z\leq s_+$ with
\begin{equation}\label{284.3}
s_-=\frac{3r_+^L+r_-^L}{2},\quad
s_+=2+\frac{(r_+^L-r_-^L)^2}{2(r_+^L+r_-^L-2)}.
\end{equation}
The wavelength in this case is given by the formula
\begin{equation}\label{si3}
  L=\frac{2\pi}{\sqrt{(\la_4(z)-r_-^L)(\la_4(z)-r_+^L)}}.
\end{equation}
Substitution of this solution into Eqs.~(\ref{273.32}) yields the
dependence of the parameters $w_i$ on $z$ which, in turn, determines --
according to Eq.~(\ref{eq33}) --- the oscillatory structure of $w(x,t)$
in a new type of shock which we shall call, as mentioned above, a
{\it contact dispersive shock wave}.

In a similar way, we may consider the diagram represented in
Fig.~\ref{fig11}(b) which corresponds in the $(v,w)$-plane to paths
crossing the anti-diagonal $w=-v$. The solution of the Whitham
equations takes the form [see Eq. \eqref{277.58}]
\begin{equation}\label{284.2}
\begin{split}
  & v_1=v_2=2\la_1+\frac{(r_+^L-r_-^L)^2}{2(r_+^L+r_-^L-2\la_1)}=z,\\
  & \la_3=r_-^L=r_-^R,\quad \la_4=r_+^L=r_+^R,
  \end{split}
\end{equation}
or
\begin{equation}\label{285.2}
\begin{split}
  \la_1(z)=\frac14 &
\bigg[z+r_+^L+r_-^L-\\
& \sqrt{(z-r_+^L-r_-^L)^2+2(r_+^L-r_-^L)^2}\bigg]
\end{split}
\end{equation}
where $z$ belongs to the interval $s_-\leq z\leq s_+$ with
\begin{equation}\label{285.3}
s_-=-2+\frac{(r_+^L-r_-^L)^2}{2(r_+^L+r_-^L+2)},\quad
  s_+=\frac{r_+^L+3r_-^L}{2}.
\end{equation}
The wavelength is here given by the formula
\begin{equation}\label{si4}
  L=\frac{2\pi}{\sqrt{(\la_1(z)-r_-^L)(\la_1(z)-r_+^L)}}.
\end{equation}

\begin{figure}
  \centering
  \includegraphics[width=9cm]{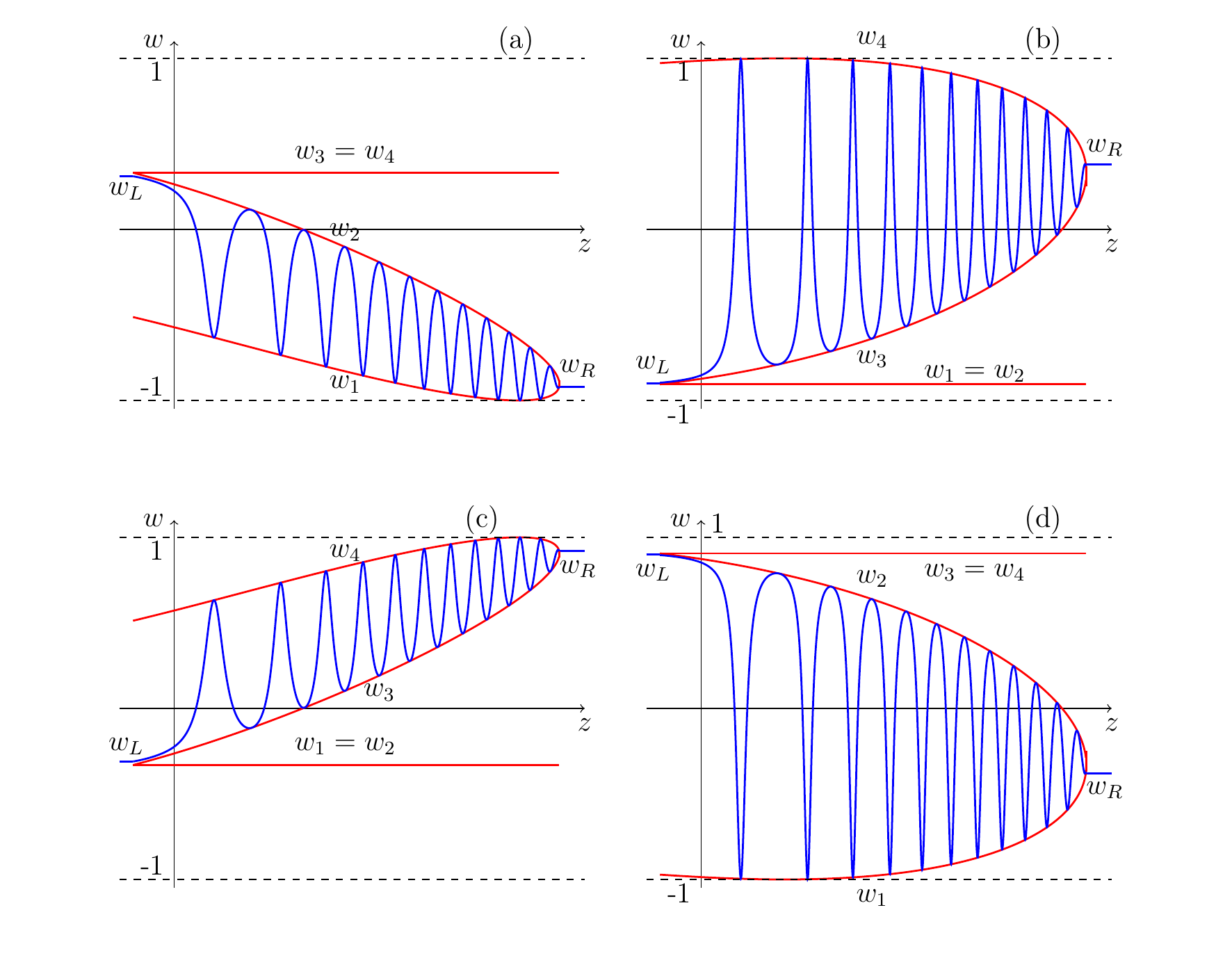}
  \caption{Contact dispersive shock waves (blue) and associated
    envelope functions $w_i$ (red) describing the solution of the
    Whitham equations which is represented by the diagram shown in
    Fig.~\ref{fig11}(a) with $r_-^L=r_-^R=-0.7$, $r_+^L=r_+^R=0.1$.
    The four shocks correspond to the four possible choices for the
    parameter $f_1$ in Eqs.~(\ref{270.16}).}
  \label{fig12}
\end{figure}
Now, as in the case of cnoidal DSWs, we have to determine how these
solutions of the Whitham equations are mapped onto the envelop
parameters $w_i$. For example, if we take $f_1>0$ given by
Eq.~(\ref{270.16a}), then in the limit $m\to0$ these parameters are
presented by the formulas (\ref{273.32}) and $w$ oscillates in the
interval $w_1\leq w\leq w_2$ leading to the trigonometric modulated
wave (\ref{eq26}). This situation is depicted in
Fig.~\ref{fig12}(a). Obviously, it corresponds to the transition
$A_1\to B_1$ in the $(v,w)$-plane. At the
soliton edge with $\la_4=\la_2$ we obtain  for the
parameters $w_i$ the following expressions:
\begin{subequations}\label{si5}
\begin{equation}\label{si5a}
  w_1=-\sqrt{\tfrac12\left[1+(2\la_2^2-1)S_{1,2}^{(+)}
-2\la_2\la_2'E_{1,2}^{(-)}\right]},
\end{equation}
\begin{equation}\label{si5b}
  w_2=-\sqrt{\tfrac12\left[1+(2\la_2^2-1)S_{1,2}^{(+)}
+2\la_2\la_2'E_{1,2}^{(-)}\right]},
\end{equation}
\end{subequations}
and at the small amplitude edge (where $\la_4=1$) the expression
\begin{equation}\label{si6}
  w_1=w_2=-\sqrt{\tfrac12(1+\la_1\la_2+\la_1'\la_2')},
\end{equation}
with the same formula (\ref{273.32c}) for $w_3$ and $w_4$ at  both
edges.

\begin{figure}
  \centering
  \includegraphics[width=8cm]{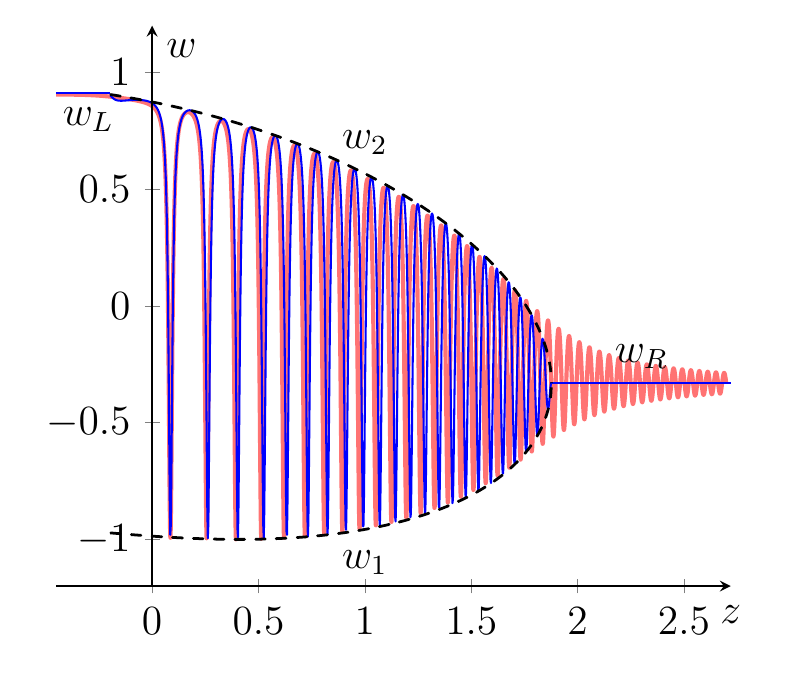}
  \caption{Comparison of the Whitham analytic solution shown in
    Fig.~\ref{fig12}(d) with the exact numerical solution (red line)
    of the Landau-Lifshitz equations after an evolution time $t=100$
    with the same boundary conditions as in Fig. \ref{fig12}(d):
    $(v_L,w_L)=(-0.33,0.91)$, $(v_R,w_R)=(0.91,-0.33)$ which
    corresponds to $r_-^L=r_-^R=-0.7$, $r_+^L=r_+^R=0.1$. }
  \label{fig13}
\end{figure}

If instead we consider the case where $f_1>0$ is given
Eq.~(\ref{270.16b}), we obtain the CDSW shown in Fig.~\ref{fig12}(b)
which corresponds to the opposite transition $B_1\to A_1$.  At the
soliton edge (where $\la_4=\la_2$) we obtain the expressions
\begin{subequations}\label{si7}
\begin{equation}\label{si7a}
  w_3=-\sqrt{\tfrac12\left[1+(2\la_2^2-1)S_{1,2}^{(-)}
+2\la_2\la_2'E_{1,2}^{(+)}\right]},
\end{equation}
\begin{equation}\label{si7b}
  w_4=\sqrt{\tfrac12\left[1+(2\la_2^2-1)S_{1,2}^{(-)}
-2\la_2\la_2'E_{1,2}^{(+)}\right]},
\end{equation}
\end{subequations}
and at the small amplitude edge (where $\la_4=1$)
\begin{equation}\label{si8}
  w_3=-w_4=-\sqrt{\tfrac12(1+\la_1\la_2-\la_1'\la_2')},
\end{equation}
with the same formula (\ref{273.36a}) for $w_1$ and $w_2$ at both
edges. Considering the other cases, with $f_1<0$ leads to CDSWs
represented in Figs.~\ref{fig12}(c,d) and corresponding to the
transitions $B_2\to A_2$ and $A_2\to B_2$ respectively. In
Fig.~\ref{fig13} we compare the analytic solution with the exact
numerical solution of the Landau-Lifshitz equation for the boundary
conditions corresponding to Fig.~\ref{fig12}(d). As we see, there is
very good agreement of the envelope functions with the numerical
results

In a similar way one can consider solutions schematically depicted in
Fig.~\ref{fig11}(b). They correspond to transitions $A_{1}
\leftrightarrow A_2$ or $B_{1} \leftrightarrow B_2$ which cross the
anti-diagonal $w=-v$.  If we take $f_1>0$ given by
Eq.~(\ref{270.16a}), at the soliton edge ($\la_2=\la_3$) we obtain the
expressions
\begin{subequations}\label{si9}
\begin{equation}\label{si9a}
  w_3=\sqrt{\tfrac12\left[1+(2\la_3^2-1)S_{3,4}^{(+)}
-2\la_3\la_3'E_{3,4}^{(-)}\right]},
\end{equation}
\begin{equation}\label{si9b}
  w_4=\sqrt{\tfrac12\left[1+(2\la_3^2-1)S_{3,4}^{(+)}
-2\la_3\la_3'E_{3,4}^{(-)}\right]},
\end{equation}
\end{subequations}
and at the small amplitude edge ($\la_2=-1$)
\begin{equation}\label{si10}
  w_3=w_4=\sqrt{\tfrac12(1+\la_3\la_4+\la_3'\la_4')},
\end{equation}
with the same formula (\ref{273.33a}) for $w_1$ and $w_2$ at both
edges. If we take $f_1<0$, then these expressions merely change
sign upon appropriate reordering. At last, for the case $f_1>0$ given
by Eq.~(\ref{270.16b}) we obtain at the soliton edge ($\la_2=\la_3$)
the expressions
\begin{subequations}\label{sik11}
\begin{equation}\label{sik11a}
  w_3=-\sqrt{\tfrac12\left[1+(2\la_3^2-1)S_{3,4}^{(-)}
-2\la_3\la_3'E_{3,4}^{(+)}\right]},
\end{equation}
\begin{equation}\label{sik11b}
  w_4=\sqrt{\tfrac12\left[1+(2\la_3^2-1)S_{3,4}^{(-)}
-2\la_3\la_3'E_{3,4}^{(+)}\right]},
\end{equation}
\end{subequations}
and at the small amplitude edge ($\la_2=-1$)
\begin{equation}\label{si12}
  w_3=-w_4=-\sqrt{\tfrac12(1+\la_3\la_4-\la_3'\la_4')},
\end{equation}
with the same formula (\ref{274.37a}) for $w_1$ and $w_2$ at both
edges.  Again, if we take $f_1<0$, then these expressions will also
change signs upon appropriate reordering.

\begin{figure}
  \centering
  \includegraphics[width=\linewidth]{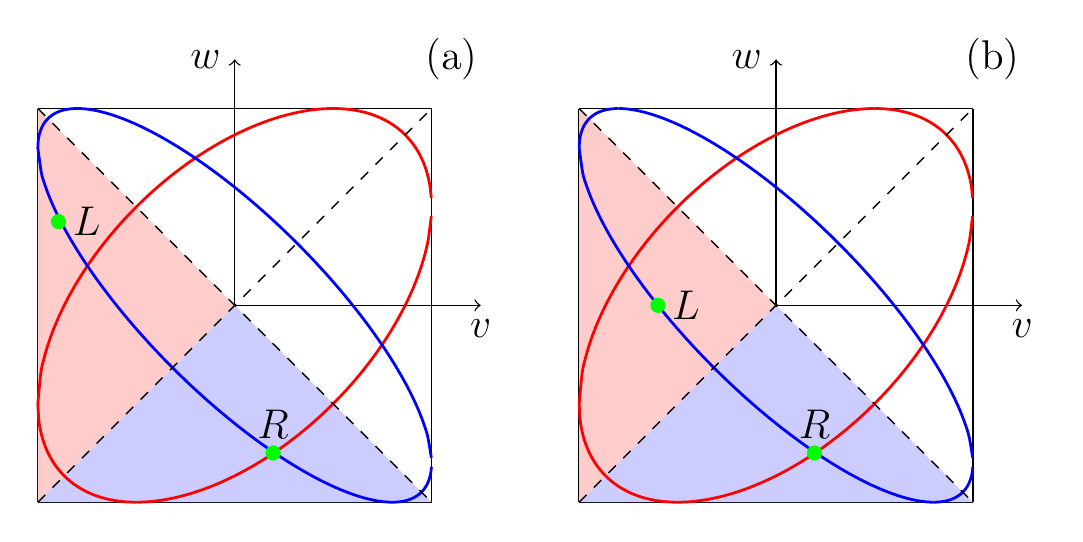}
  \caption{Paths in the $(v,w)$-plane associated with two types of
    combined shocks. The left and right boundary conditions correspond
    to points $L$ and $R$ respectively; they lie on an ellipse along
    which the dispersionless Riemann invariant $r_-$ ($=r_-^L=r_-^R$) is
    constant. One has $r_+^L<r_+^R$ in case (a) and $r_+^L>r_+^R$ in
    case (b).}
  \label{fig14}
\end{figure}
\begin{figure}
  \centering
  \includegraphics[width=\linewidth]{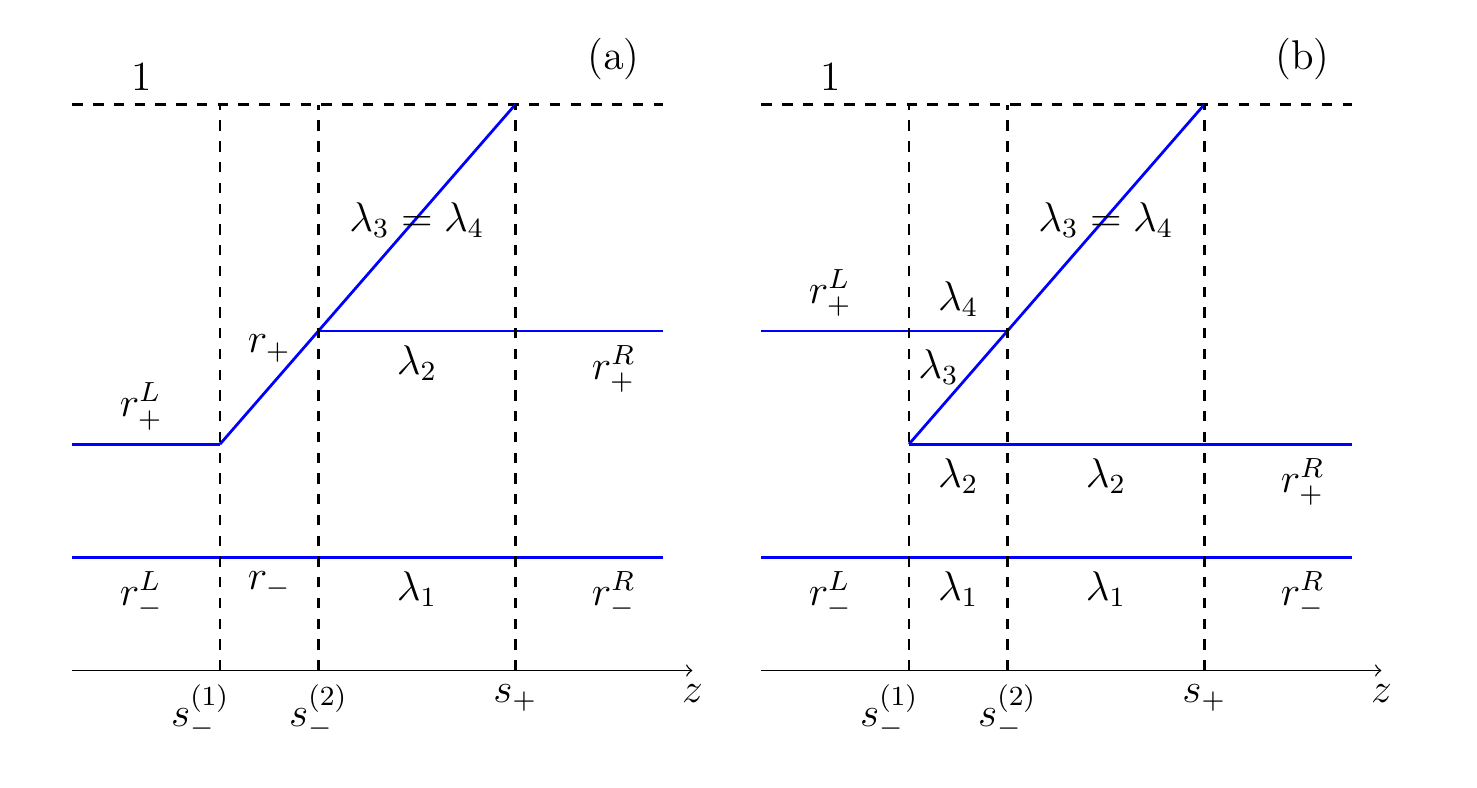}
  \caption{Sketches of the behavior of the Riemann invariants
    corresponding to the transitions in the $(v,w)$-plane shown in
    Fig.~\ref{fig14}.}
  \label{fig15}
\end{figure}

We now turn to the study of the generalizations of CDSWs by
considering the transitions depicted in Fig.~\ref{fig14}: in these
cases the boundary points are also not in the same monotonicity
triangle of the $(v,w)$ plane, still on the same ellipse because the
left and right boundary conditions have a common value for one of the
Riemann invariants (say, $r_-^L=r_-^R$), however the boundary values
of the other Riemann invariants are different ($r_+^L\neq r_+^R$). To
be definite, we shall consider two generalizations of the situation
leading to the CDSW represented in Figs.~\ref{fig12}(d) and
\ref{fig13}. The transition of the type $A_2\to B_2$ of
Fig.~\ref{fig10} can be generalized in two ways represented in
Fig.~\ref{fig14}, where the points $L$ and $R$ symbolize plateaus at
the left and right boundaries, respectively. In this case
$r_-^L=r_-^R$ because the transition occurs along the ellipse where
this Riemann invariant is constant. As we know, the dispersionless
invariant $r_+$ decreases along such a curve when going away from the
diagonal $w=v$, hence we have $r_+^L<r_+^R$ and $r_+^L>r_+^R$ in case
(a) and (b), respectively. This suggests the generalizations of the
diagram Fig.~\ref{fig11}(a) depicted in Fig.~\ref{fig15}.

\begin{figure}
  \centering
  \includegraphics[width=0.9\linewidth]{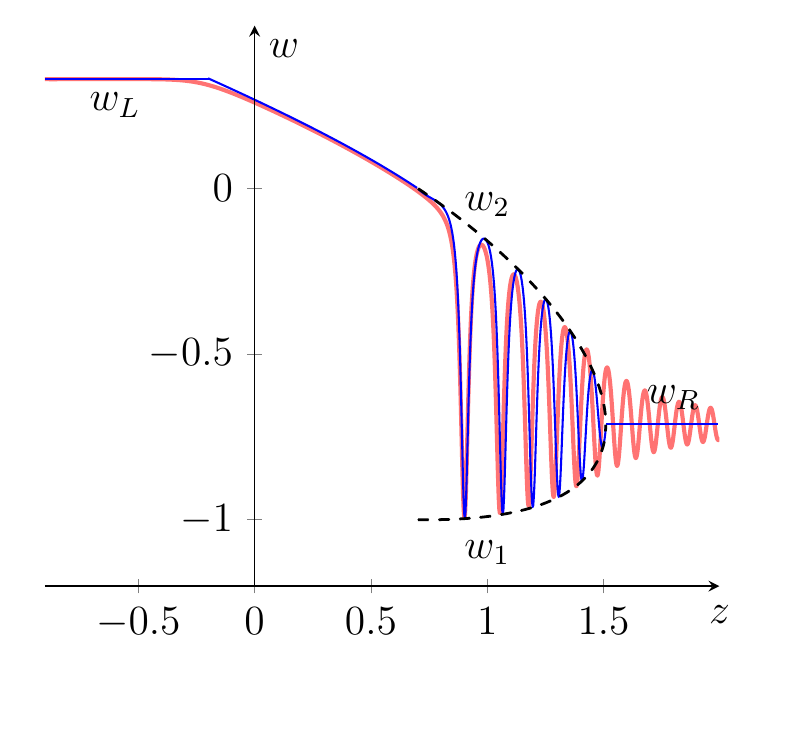}
  \caption{Comparison of the analytic solution corresponding to the
    left and right boundary conditions depicted in Fig.~\ref{fig15}(a)
    with the exact numerical solution of the Landau-Lifshitz
    equations.}
  \label{fig16}
\end{figure}

In the case corresponding to Fig.~\ref{fig15}(a), the CDSW is attached
at its soliton edge to a rarefaction wave which matches at its left
edge with the left boundary plateau. The velocities of the
characteristic points identified in Fig.~\ref{fig15}(a)
are expressed in terms of the boundary
Riemann invariants by the formulas
\begin{equation}\label{si13}
  \begin{split}
  &s_-^{(1)}=\frac12(r_-^L+3r_+^L), \,\,\,\, s_-^{(2)}=\frac12(3r_+^R+r_-^R),\\
  &s_+=2+\frac{(r_+^R-r_-^R)^2}{2(r_+^R+r_-^R-2)}.
  \end{split}
\end{equation}
The resulting composite wave structure is shown in Fig.~\ref{fig16}
(blue line) where it is compared with the numerical solution of the
Landau-Lifshitz equation (red line).

\begin{figure}
  \centering
  \includegraphics[width=0.9\linewidth]{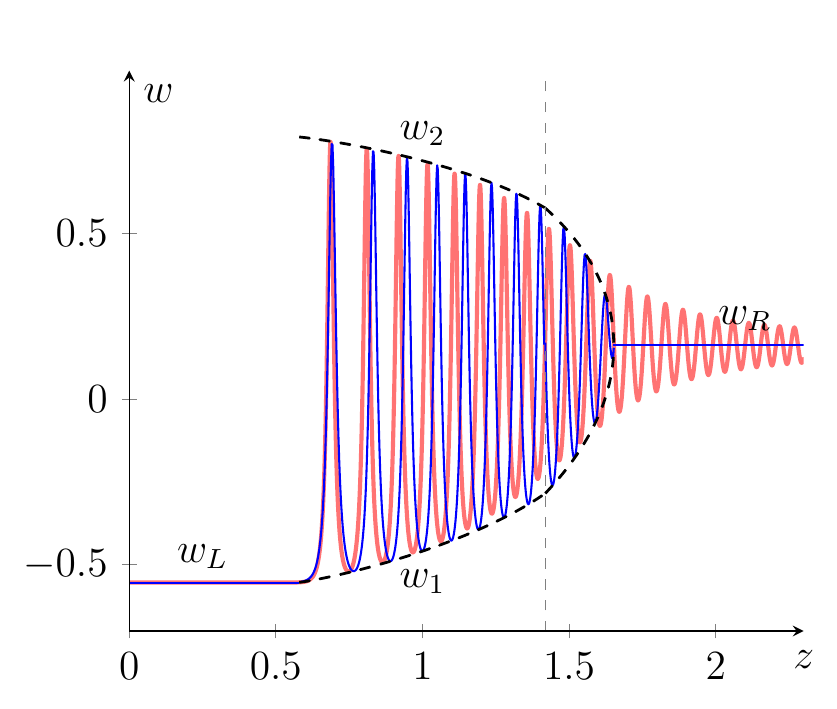}
  \caption{Comparison of the analytic solution corresponding to the
    left and right boundary conditions depicted in Fig.~\ref{fig15}(b)
    with the exact numerical solution of the Landau-Lifshitz
    equations. The (analytically determined) vertical dashed line
    separates the cnoidal wave (at the left) from the trigonometric
    wave (at the right).}
  \label{fig17}
\end{figure}

In the case corresponding to Fig.~\ref{fig15}(b) the trigonometric
CDSW is attached to a cnoidal DSW of the type Fig.~\ref{fig8}(a) which
degenerates at its right edge (at which $w_2=w_1$) into a trigonometric
wave. At the left soliton edge the cnoidal wave matches with the left
boundary plateau. The velocities of the characteristic points
identified in Fig.~\ref{fig15}(b) are given by
\begin{equation}\label{si14}
  \begin{split}
  &s_-^{(1)}=\frac{1}{2}(r_-^L+2r_+^R+r_+^L), \\
& s_-^{(2)}=2r_+^L+\frac{(r_+^R-r_-^R)^2}{2(r_+^R+r_-^R-2r_+^L)},\\
  &s_+=2+\frac{(r_+^R-r_-^R)^2}{2(r_+^R+r_-^R-2)}.
  \end{split}
\end{equation}
The resulting composite wave structure is shown in Fig.~\ref{fig17}
(blue line) where it is compared with the numerical solution of the
Landau-Lifshitz equation (red line).

It is clear that any transition between points shown in
Fig.~\ref{fig10} can be generalized in a similar way leading to
composite shock waves consisting of cnoidal, trigonometric and
rarefaction waves. We shall not list here all these possible wave
structures since the general principles for their construction are
simply deduced from the examples just presented.

This ends the characterization of all the key elements which may
appear in a complex wave structure evolving from an arbitrary initial
discontinuity of type \eqref{init-cond}. We can now proceed to the
classification of all the possible composite structures.

\section{Classification}\label{sec5}

As clear from the previous section, it is convenient to distinguish
the situations where both points representing the left and right
boundary conditions belong to the same triangle of monotonicity from
those where they belong to different such triangles. It has been
noticed in subsection \ref{NLS0} that in these triangles, for some
limiting values of the variables, the Landau-Lifshitz equation reduces
either to the NLS or to KB equation. We shall thus refer to such
triangles as being of ``NLS type'' or of ``KB type'', and consider
them separately.

\subsection{Nonlinear Schr\"odinger type sector}\label{sec5a}

\begin{figure}
  \centering
  \includegraphics[width=\linewidth]{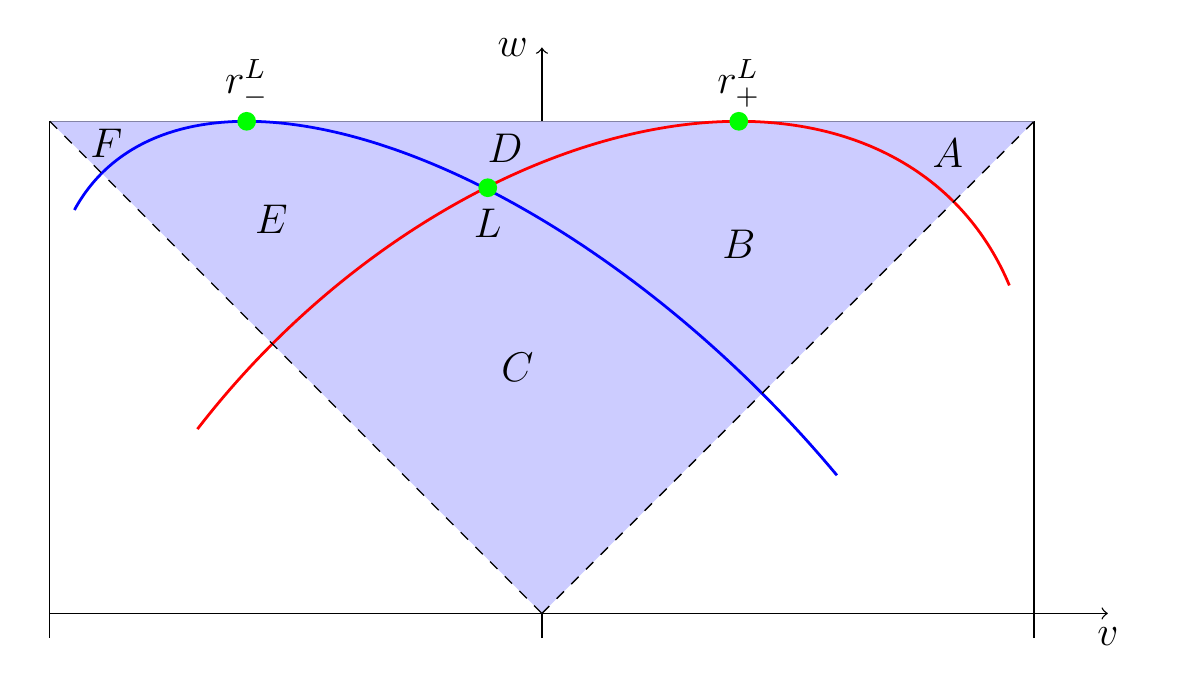}
  \caption{Plot of the upper monotonicity triangle of NLS type in the
    $(v,w)$ plane. The (red and blue) curves of constant
    dispersionless Riemann invariants $r_{\pm}^L$ corresponding to the
    left boundary point $L$ divide this triangle into six domains
    denoted as A, B, $\ldots$ , F. The type of flow depends on the
    domain in which lies the right boundary point $R$ of coordinates
    $(v_R,w_R)$.}
  \label{fig18}
\end{figure}

Since the theory for the upper and lower NLS type triangles is
essentially the same, we shall confine ourselves to the
upper triangle which is shown in Fig.~\ref{fig18}.

\begin{figure}
  \centering
  \includegraphics[width=4cm]{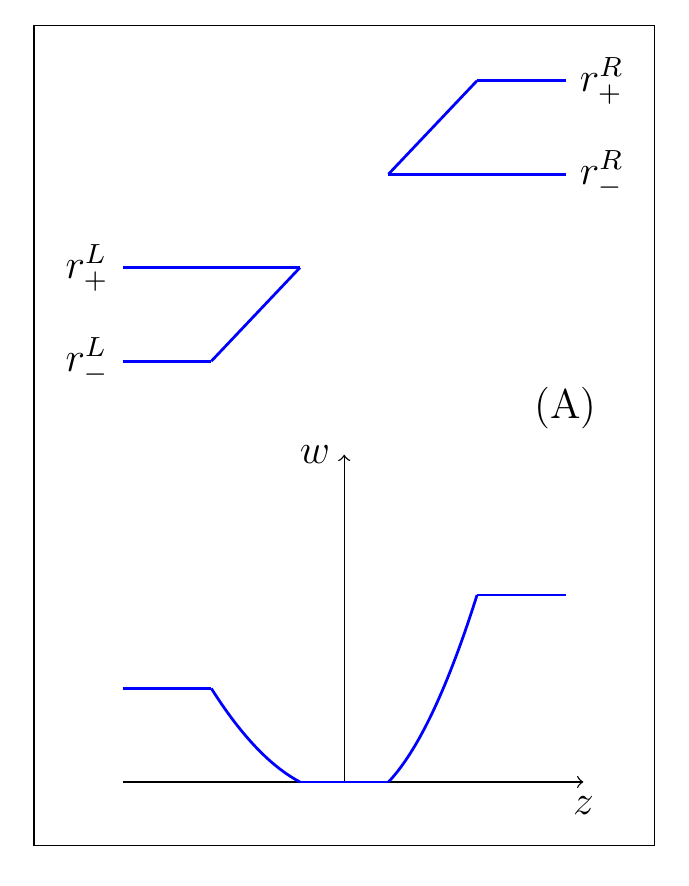}
  \includegraphics[width=4cm]{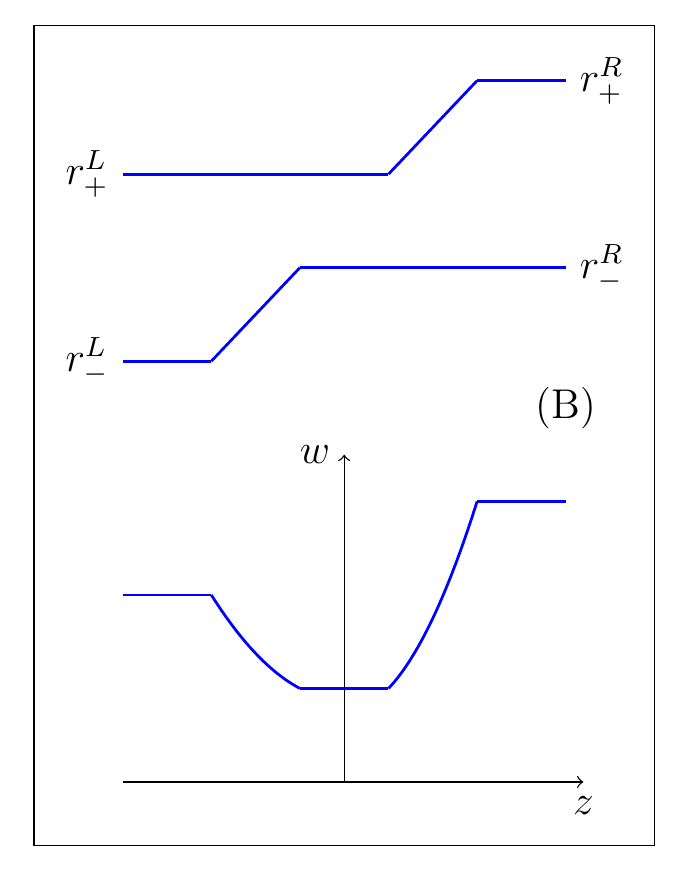}\\
  \includegraphics[width=4cm]{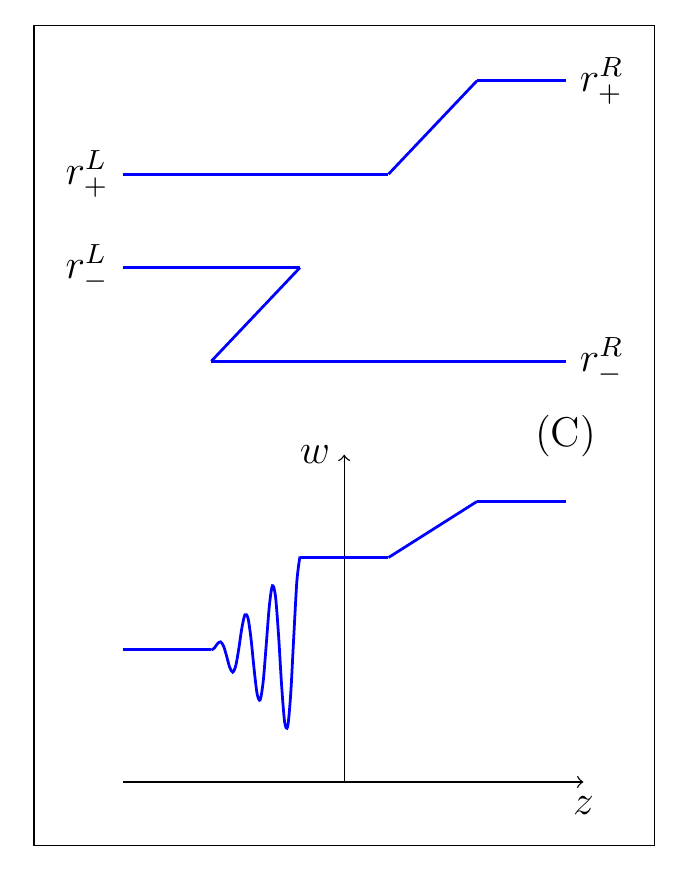}
  \includegraphics[width=4cm]{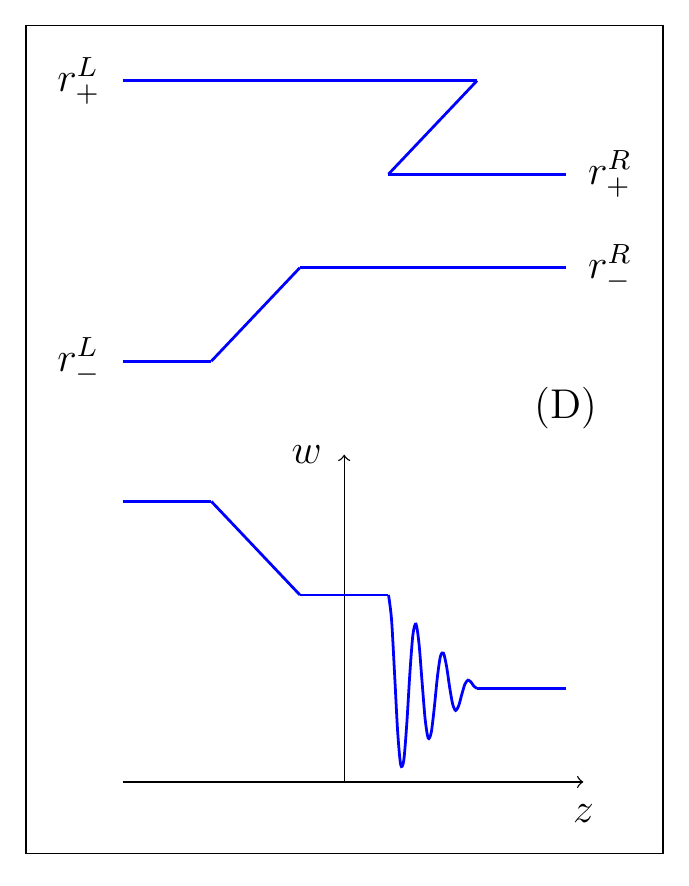}\\
  \includegraphics[width=4cm]{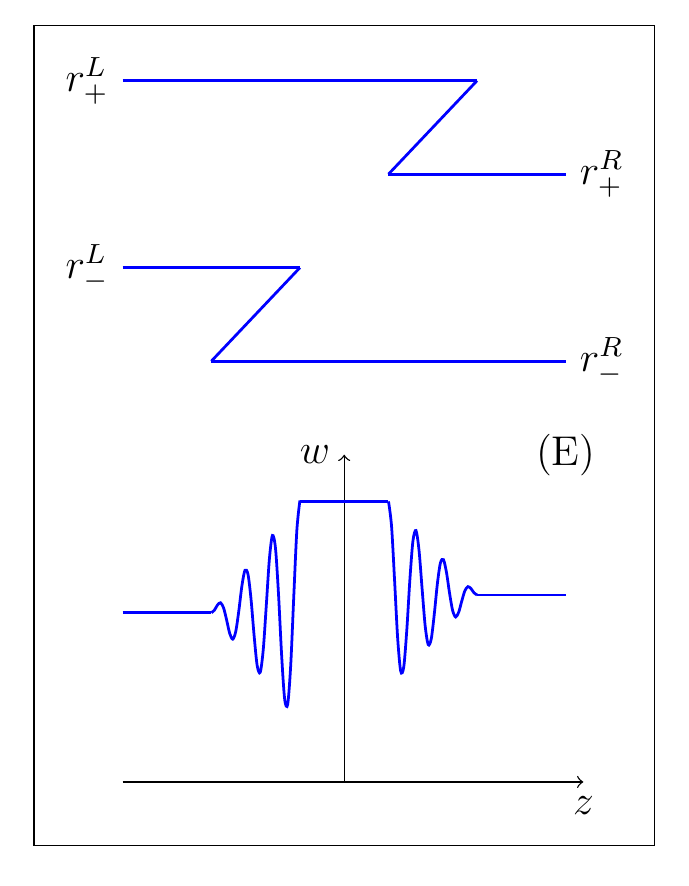}
  \includegraphics[width=4cm]{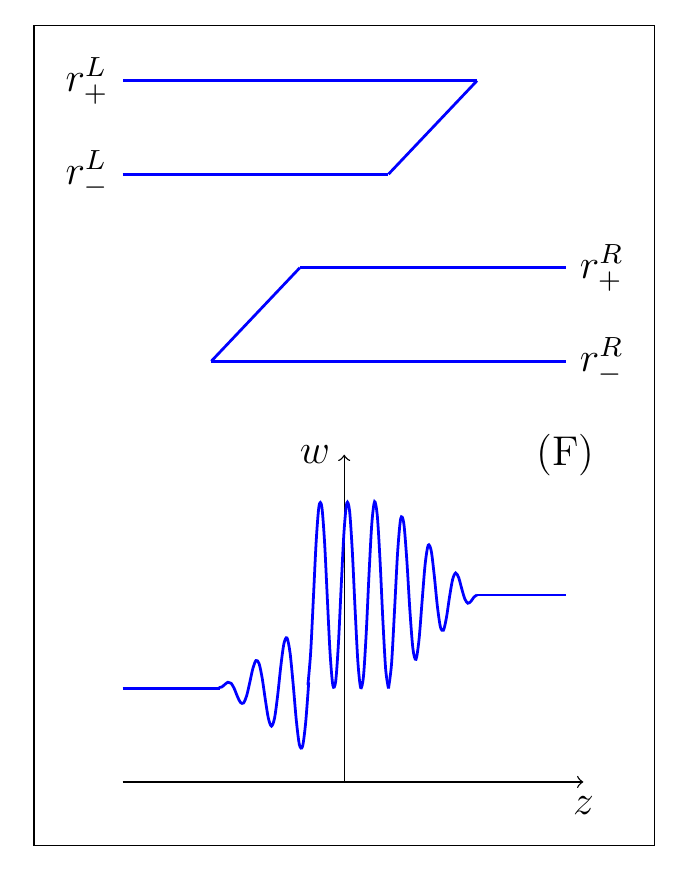}
  \caption{Sketches of the behavior of the Riemann invariants and of
    the corresponding wave structures corresponding to the location of
    the point $R$ referring to the right boundary in one of the six
    domains shown in Fig.~\ref{fig18}.}
  \label{fig19}
\end{figure}
We thus consider the case where both left and right initial conditions
correspond to points located inside this triangle of the
$(v,w)$-plane. For definiteness, we denote the point of coordinates
$(v_L,w_L)$ referring to the left boundary by $L$ and plot the two
ellipses of constant Riemann invariants $r_{+}^L$ and $r_{-}^L$.
These divide the triangle into six sub-domains. It is easy to see
that, when the point $R$ referring to the right boundary is located in
one of these domains (labelled by the symbols A, B, ..., F), one of
the following inequalities is fulfilled:
\begin{equation}\label{290.1}
  \begin{split}
  & ({\rm A})\quad r_-^L < r_+^L < r_-^R < r_+^R,\quad
    ({\rm B})\quad r_-^L < r_-^R < r_+^L < r_+^R,\\
  & ({\rm C})\quad r_-^R < r_-^L < r_+^L < r_+^R,\quad
    ({\rm D})\quad r_-^L < r_-^R < r_+^R < r_+^L,\\
  & ({\rm E})\quad r_-^R < r_-^L < r_+^R < r_+^L,\quad
    ({\rm F})\quad r_-^R < r_+^R < r_-^L < r_+^L.
  \end{split}
\end{equation}
The corresponding diagrams of the Riemann invariants symbolizing the
self-similar solutions of the Whitham equations and sketches of wave
structures are shown in Fig.~\ref{fig19}.  In case (A) the structure
consists of two rarefaction waves expanding into `vacuum' and in case
(B) these two rarefaction waves are connected by a plateau whose
parameters are determined by the dispersionless Riemann invariants
$r_{\pm}^P$ equal to $r_-^P=r_-^R$ and $r_+^P=r_+^L$.  In cases (C)
and (D) the structure consists of one DSW and one rarefaction wave
connected by a plateau characterized by the same parameters. In case
(E) there are two DSWs connected by a plateau and, at last, in case
(F) the previous plateau is replaced by a nonlinear wave which --- with
high enough accuracy --- can be presented as a non-modulated cnoidal
wave. Not surprisingly, this classification coincides qualitatively
with the one obtained in Ref.~\cite{El95} for the NLS equation. It is
clear that it is determined by the geometry of the curves of constant
Riemann invariants: the arcs of ellipses shown in Fig.~\ref{fig18}
become, in the NLS equation, arcs of parabolas with the same
subdivision of the monotonicity region which, in the NLS case, extends
to the whole half-plane of all possible values of the physical
parameters. In the present Landau-Lifshitz case, a typical example of
such a structure has been studied in some details in
Ref.~\cite{Con16}.

It is important to notice that the domains A and F cannot be reached
from point $L$ in Fig. \ref{fig18} via paths consisting of arcs of
constant dispersionless Riemann invariants without by-passing the
points labelled as $r_-^L$ and $r_+^L$, at which the meaning of the
Riemann invariants changes (see Fig. \ref{fig2} and the related
discussion in the text). Therefore, in these two cases, the edge wave
structures are separated either by vacuum (i.e. $w=0$) in case (A), or
by a cnoidal wave in case (F). In the other situations (B, C, D, E) we
can draw two arcs of constant invariant ellipses whose crossing point
defines the plateau connecting the edge wave structures (rarefaction
waves or DSWs). It is easy to see that these two arcs can be drawn in
two ways and that the physically relevant one is distinguished by the
condition that the speeds of the matching points increase from left to
right (see a similar argumentation in the theory of standard viscous
shock waves in Ref. \cite{LL-6}). Consequently, the left edge wave
always corresponds to a diagram of Riemann invariants with
$r_+^L=\mathrm{const}$ continued through the whole left wave, whereas
in the right edge wave we have $r_-^R=\mathrm{const}$ also continued
through the whole right structure. The diagrams (B) to (E) in
Fig.~\ref{fig19} illustrate this simple principle. This remark removes
any ambiguity in the determination of the wave structure arising from
initial conditions referring to the NLS type sectors.

\subsection{Kaup-Boussinesq type sector}\label{sec5b}

\begin{figure}
  \centering
  \includegraphics[height=0.8\linewidth]{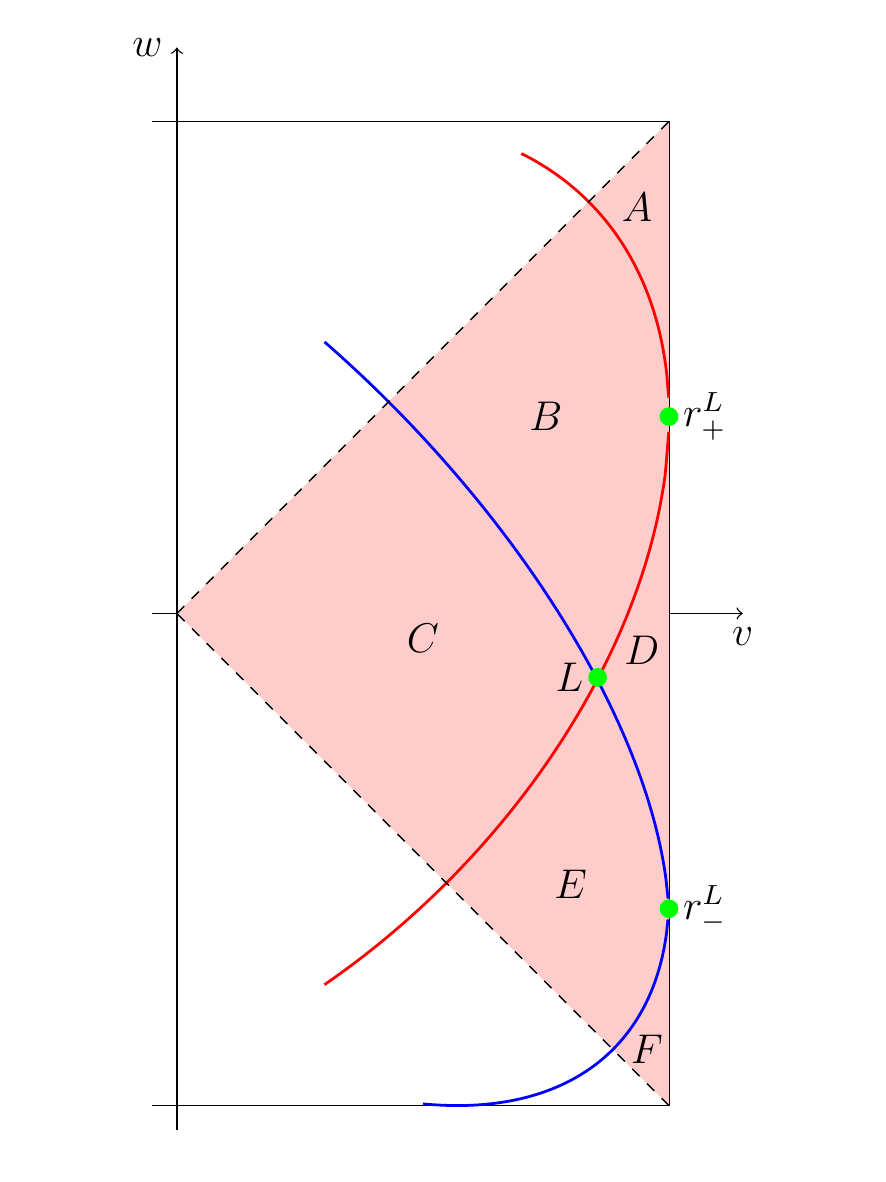}
  \caption{Plot of the right monotonicity triangle of KB type in the
    $(v,w)$ plane. The (red and blue) curves of constant
    dispersionless Riemann invariants $r_{\pm}^L$ corresponding to the
    left boundary point $L$ divide this triangle into six domains
    denoted as A, B, $\ldots$ , F. The type of flow depends on the
    domain in which lies the right boundary point $R$ of coordinates
    $(v_R,w_R)$.}
  \label{fig20}
\end{figure}

We shall now consider initial conditions for which both left and right
boundary points are located in a KB sector of the $(v,w)$-plane which
consists in one of the triangles delimited by the diagonal, the
anti-diagonal and the vertical curve $v=\pm 1$. For definiteness we
consider the right KB triangle. This situation bares many similarities
with the preceding NLS type case. Indeed, from Fig.~\ref{fig20} we see
that, again, the monotonicity triangle is divided into six domains
corresponding to the inequalities listed in Eq.~(\ref{290.1}) --- these
domains are symmetrical with respect to the diagonal $w=v$ to those
shown in Fig.~\ref{fig18}. It is clear that the diagrams of Riemann
invariants and the corresponding wave structures are qualitatively the
same as the ones depicted in Fig.~\ref{fig19}. A detailed discussion
of the Riemann problem for the KB equation \eqref{KB-canonical} has
been recently given in Ref.~\cite{KB17} and in the KB sector of
Landau-Lifshitz equation theory the resulting wave patterns are
qualitatively the same --- they consist of DSWs and/or rarefaction
waves connected with each others by plateaus.

As in the NLS type sectors, if the point $R$ referring to the right
boundary lies in one of the domains B, C, D or E, it can be connected
with $L$ by two arcs of constant Riemann invariant ellipses in two
possible ways; the physically acceptable one is identified by the
condition that the speeds of the matching points increase from left to
right. The crossing point of these two arcs defines the parameter of
the plateau which connects the two edge wave structures. In cases (A)
and (F) the plateau does not exist and is replaced either by
a vacuum region or by a non-modulated cnoidal wave. We thus arrive at
the same wave structures that the ones illustrated in Fig.~\ref{fig19}.

\subsection{Wave structures with transitions between
monotonicity sectors}\label{sec5c}

The above formulated principles of construction of diagrams for the
Riemann invariants make it possible to predict which wave structure
will evolve from a given boundary condition (\ref{init-cond}), even in
cases where the left and right boundary points belong to different
triangles of monotonicity. Since the total number of possible wave
structures is very large, we shall not list all of them here but
rather illustrate the principles of construction by an application to
a typical particular case.

\begin{figure}
  \centering
  \includegraphics[width=\linewidth]{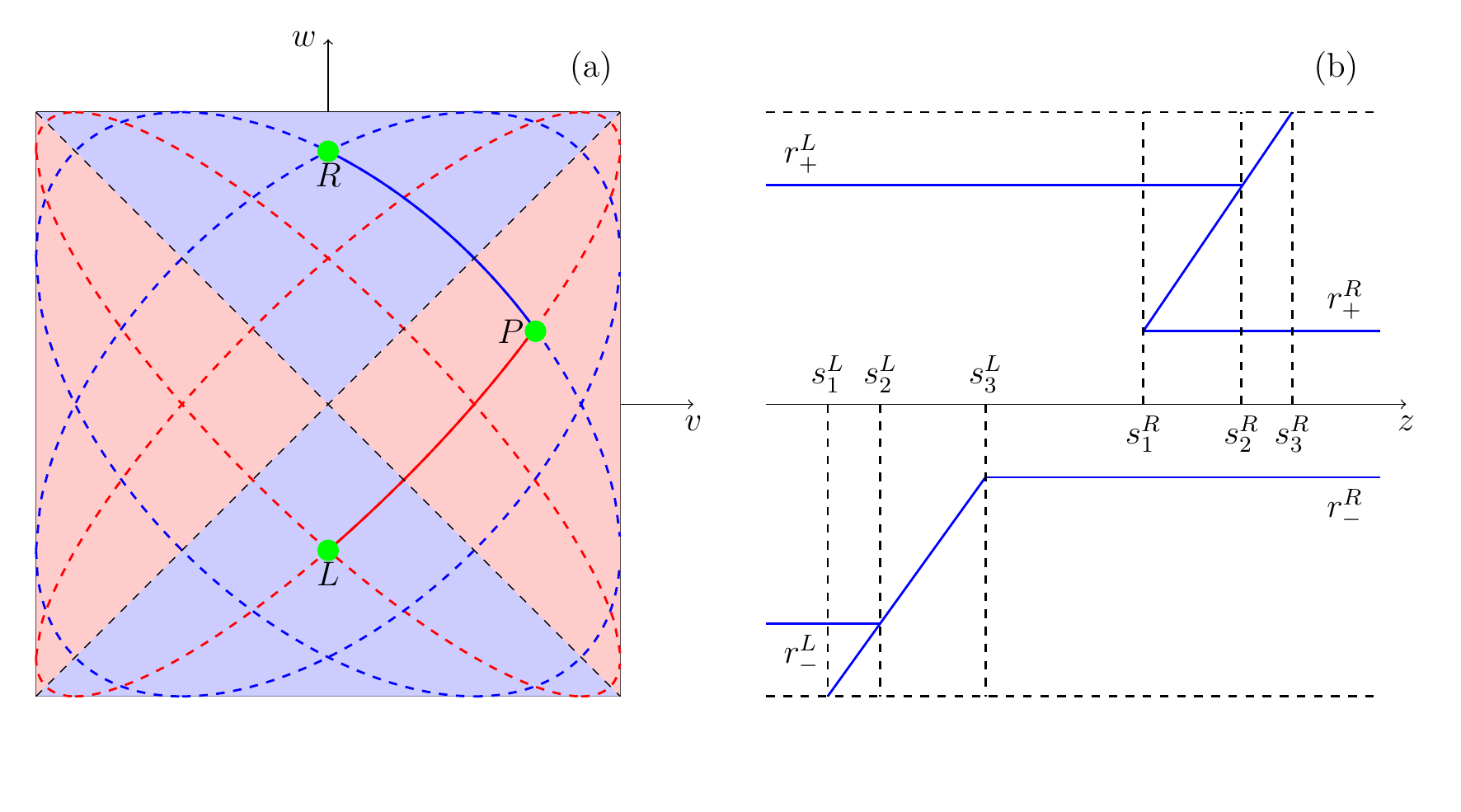}
  \caption{(a) Plot in the $(v,w)$ plane of ellipses along which the
    dispersionless Riemann invariants are constant. They are
    represented by red (blue) dashed lines for the left (right)
    boundary conditions.  The path connecting the left $L$ and the
    right $R$ points are shown by solid lines. They intersect at point
    $P$ representing the plateau located between the left and right
    waves.  (b) Sketch of the behavior of the Riemann invariants
    corresponding to solutions of the Whitham equations for the same
    boundary conditions.  The left wave consists of a trigonometric
    shock (for $s_1^L<z<s_2^L$) attached to a rarefaction wave (for
    $s_2^L<z<s_3^L$). The right wave is a combined cnoidal
    ($s_1^R<z<s_2^R$) and trigonometric ($s_3^R<z<s_3^R$) shock.}
  \label{fig21}
\end{figure}

Let us take $v_L=v_R=0$, $w_L<0$, $w_R>0$ and $|w_L|<w_R$. We see at
once from Fig.~\ref{fig21}(a) that the dispersionless ellipses
relating $L$ to $R$ must cross both diagonals of the hyperbolicity
square, hence the wave structure must consist of two contact and/or
combined waves. Substitution of the above parameters into
Eq.~(\ref{nd2}) yields the values of the dispersionless Riemann
invariants which are ordered according to
$r_-^L<r_-^R<r_+^R<r_+^L$. Taking into account that the left wave
corresponds to the continuation of $r_+^L=\mathrm{const}$ and the
right wave to the continuation of $r_-^R=\mathrm{const}$, we arrive at
the diagram shown in Fig.~\ref{fig21}(b). At the left edge we have the
combination of a trigonometric shock ($s_1^L\leq z\leq s_2^L$) with a
rarefaction wave ($s_2^L\leq z\leq s_3^L$) and, at the right edge, one has
merged cnoidal ($s_1^R\leq z\leq s_2^R$) and trigonometric ($s_2^R\leq
z\leq s_3^R$) shocks. These left and right edge waves are connected
one with the other by a plateau characterized by the Riemann
invariants $r_-^P=r_-^R$ and $r_+^P=r_+^L$. This plateau is represented by
the single point $P$ in Fig.~\ref{fig21}(a).

\begin{figure}
  \centering
  \includegraphics[width=0.9\linewidth]{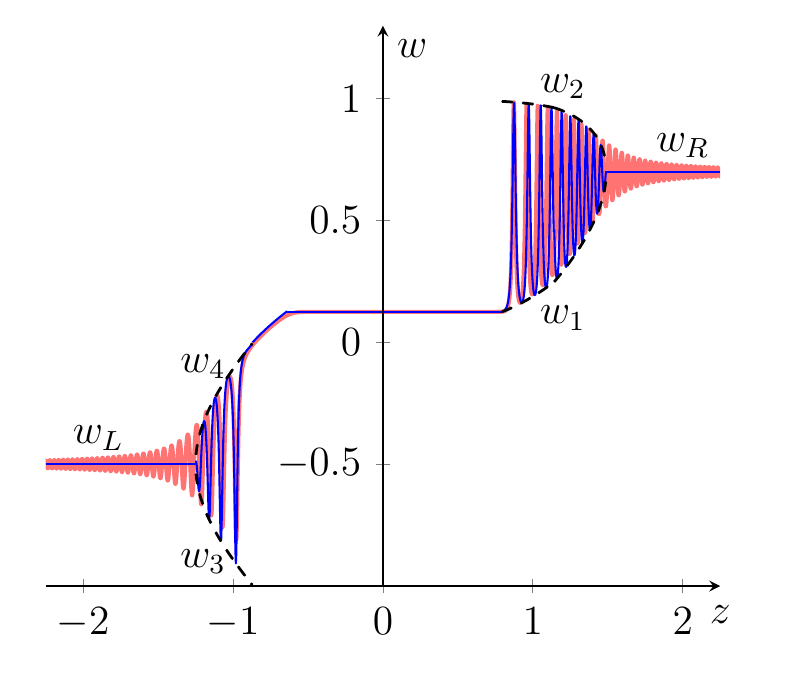}
  \caption{Comparison of analytic (blue line) and numerical (red line)
solutions for the
    initial profile \eqref{init-cond} with $v_L=v_R=0$, $w_L=-0.5$,
    $w_R=0.7$. The left and right boundary points are represented in
    Fig. \ref{fig21}(a) and the behavior of the Riemann invariants is
    sketched in Fig. \ref{fig21}(b).}
  \label{fig22}
\end{figure}

The formulas connecting the zeroes $w_i$ of the resolvent with the
Riemann invariants $\la_i$ (obtained as solutions of the Whitham
equations) are of the type discussed in Section~\ref{sec4c}. They make
it possible to explicitly construct the Whitham wave structure (shown in
Fig.~\ref{fig22} by a blue line) which compares very well with the
numerical solution of the Landau-Lifshitz equation (red line).
This example clearly illustrates how the wave structure can be
constructed for any choice of parameters of the initial
discontinuous profile \eqref{init-cond}.

\section{Discussion and conclusion}\label{sec.conclu}

In this paper we have solved the Riemann problem and characterized the
space-time evolution of an initial discontinuity for the
Landau-Lifshitz equation. This equation describes magnetization
excitations in a dissipationless easy-plane ferromagnet and, in the
appropriate regime, polarization waves in a two-component BEC. It is
natural to suppose that the method developed in the present work is
general enough and should apply to many other models. We shall thus
formulate the most important points of our approach.

(i) The first obvious feature is the statement that the method in its
present form applies to modulationally stable situations only, so
the region of hyperbolicity of the long wave (dispersionless)
approximation should be determined and the boundary conditions at both
sides of the discontinuity must lie within this region.

(ii) The hyperbolicity region should be subdivided into domains where
the dispersionless approximation is {\it genuinely nonlinear} (see,
e.g., Ref. \cite{lax}), i.e., where the characteristic velocities depend on
the field variables with non vanishing gradients. In our case --- with
two field variables with known Riemann invariants --- we have denoted
such domains as {\it monotonicity sectors}. For systems described by a
single field variable, this condition reduces to imposing a fixed
convexity to the dependence of the dispersionless velocity on the
amplitude of the wave; an example of such a situation was considered
in Ref.~\cite{Kam12}.

(iii) If both boundary conditions belong to the same monotonicity
sector, then the classification of the wave structures follows
closely well-known examples such as KdV (one field variable,
\cite{gp-73}) or NLS (two field variables \cite{El95}) theories.
These wave structures consist of rarefaction waves and standard
dispersive shock waves of Gurevich-Pitaevskii type connected with each
other by a plateau, a ``vacuum'' or a two-phase (i.e., non-modulated
``cnoidal'') wave region.

(iv) If the boundary conditions belong to different monotonicity
sectors, then they are connected by profiles consisting of new
wave structures --- contact (trigonometric) dispersive shocks or
kinks. In situations with a single field variables these were
identified, respectively, in Refs.~\cite{marchant} and \cite{ep-11};
both structures appeared also in the theory of the Gardner equation
\cite{Kam12}. In the case considered here of the Landau-Lifshitz equation
we have dealt with contact dispersive shock waves and their
combinations with other structures.

(v) When the evolution equations are completely integrable, the
Whitham system can be transformed into a diagonal Riemann form and in
this case the mapping of the Riemann invariants to the physical
parameters is not single-valued. Instead, it is realized by sets of
relationships between the zeroes of two polynomials: the polynomial
whose roots are the Riemann invariants (noted $P$ in the main text)
and its algebraic resolvent (${\cal R}$). These relationships appear
in a natural way in the finite-gap integration method (see, e.g.,
Ref. \cite{ZMNP}) complemented by resolving the problem of ``reality
conditions'' \cite{kamch-90} (see also Ref. \cite{Kam00}). For the
Landau-Lifshitz equation the corresponding resolvent was found in
Ref.~\cite{Kam92} in which, however, the consequences of the
multiplicity of relationships between the Riemann invariants and
the zeroes of the resolvent were not completely elucidated. The theory
developed in the present work clarifies this important point.

We thus believe that the solution of the Riemann problem for the case
of nonlinear waves whose evolution is governed by the Landau-Lifshitz
equations provides a general scheme which applies to other systems
which share the similar characteristic property of not being genuinely
nonlinear, cf. the case of the modified NLS equation considered in
Ref. \cite{Iva17b}. Besides, the different situations considered in
the present work can find applications for describing nonlinear waves
in concrete physical situations such as ferromagnets and two-component
Bose-Einstein condensates.

\begin{acknowledgments}
  We thank M. Hoefer for fruitful exchanges and discussions and
  E. Iacocca for comments on the manuscript. AMK thanks Laboratoire de
  Physique Th\'eorique et Mod\`eles Statistiques (Universit\'e
  Paris-Saclay) where this work was started, for kind hospitality.
  This work was supported by the French ANR under grant
  ANR-15-CE30-0017 (Haralab project).
\end{acknowledgments}

\end{document}